\documentclass[fleqn,10pt]{wlscirep}
\usepackage[utf8]{inputenc}
\usepackage[T1]{fontenc}
\usepackage{lineno}
\usepackage{amsmath}
\usepackage{subcaption}
\usepackage{graphicx}

\usepackage[format=plain,justification=justified,singlelinecheck=false,font={stretch=1.125,small,sf},labelfont=bf,labelsep=space]{caption}

\usepackage{chemformula} 
\usepackage{braket}
\newcommand{\DefineAuthor}[2]{%
  \expandafter\newcommand\csname #1note\endcsname[1]{%
    \textbf{\textcolor{#2}{\textbf{#1:} ##1}}}%
  \expandafter\newcommand\csname #1\endcsname[1]{
    \textbf{\textcolor{#2}{##1}}}
  \expandafter\newcommand\csname #1cancel\endcsname[1]{%
    \textbf{\textcolor{#2}{\sout{##1}}}}%
  \expandafter\newcommand\csname #1change\endcsname[2]{%
    \textbf{\textcolor{#2}{\sout{##1} ##2}}}%
  \newenvironment{#1text}{\color{#2}}{\color{black}}
}
\definecolor{dartmouthgreen}{rgb}{0.05, 0.5, 0.06}
\DefineAuthor{LMS}{red}
\DefineAuthor{AF}{brown}
\DefineAuthor{AS}{blue}

\usepackage{xcolor}

\usepackage{algpseudocode}%
\usepackage{listings}%
\usepackage{siunitx}%
\DeclareSIUnit\electronvolt{eV}
\DeclareSIUnit\meva{\milli \electronvolt \per \angstrom}
\usepackage{textgreek}%
\usepackage{multirow}

\title{Toward the Rational Design of Molecular Field-Coupled Nanocomputing Candidates}

\author[1,*]{Federico Ravera}
\author[2,*]{Leonardo Medrano Sandonas}
\author[3]{Andrea Vezzoli}
\author[1]{Yuri Ardesi}
\author[4]{Mariagrazia Graziano}
\author[1]{Gianluca Piccinini}
\author[2,5,6,*]{Gianaurelio Cuniberti}

\affil[1]{Department of Electronics and Telecommunication, Politecnico di Torino, Torino, Italy}
\affil[2]{Institute for Materials Science and Max Bergmann Center of Biomaterials, TUD Dresden University of Technology, 01062, Dresden, Germany}
\affil[3]{Department of Chemistry, University of Liverpool, Liverpool, United Kingdom}
\affil[4]{Department of Applied Science and Technology, Politecnico di Torino, Torino, Italy}
\affil[5]{Cluster of Excellence CARE, TU Dresden and RWTH Aachen, Germany}
\affil[6]{Cluster of Excellence CeTI, TU Dresden, Germany}
\affil[*]{Corresponding authors:  Federico Ravera (federico.ravera@polito.it), Leonardo Medrano Sandonas (leonardo.medrano@tu-dresden.de), Gianaurelio Cuniberti (gianaurelio.cuniberti@tu-dresden.de)}

\begin{abstract}
Molecular Field-Coupled Nanocomputing (MolFCN) is a promising beyond-CMOS paradigm in which information is propagated electrostatically rather than through charge transport, enabling ultra-low-power logic. Identifying molecules with stable logic states, efficient clock-field switching, and reliable information propagation, however, remains an open challenge.
In this Letter, we introduce LUFFY (Layered Unified Framework for MolFCN systematic analYsis), a framework for the rational design and validation of molecular candidates for MolFCN architectures. Starting from 27 synthetically accessible molecules, we combine conformational sampling and electrostatic analysis in neutral and oxidized states to derive robust descriptors of molecular response. In particular, we extract the V${in}$-to-Aggregated-Charge Transcharacteristics (VACTs), capturing the field-induced charge response, and introduce energy-averaged models validated via ab initio molecular dynamics to account for conformational diversity.
Finally, we use the resulting molecular responses to evaluate device-level propagation and demonstrate stable information transfer. These results directly link molecular structure to functional information flow, identifying conformationally robust electrostatic response as a key requirement for MolFCN operation.
Overall, this work establishes a unified and transferable framework for the identification and validation of MolFCN molecular candidates, bridging molecular design and circuit-level functionality. By unifying previously fragmented approaches into a sustainable methodology, LUFFY enables rational and scalable molecular discovery and establishes a foundation for data-driven design strategies that accelerate the development of ultra-low-power information processing technologies.
\end{abstract}

\begin{document}

\flushbottom
\maketitle

\thispagestyle{empty}

\begin{figure}[t!]
    \centering
    \includegraphics[width=0.8\linewidth]{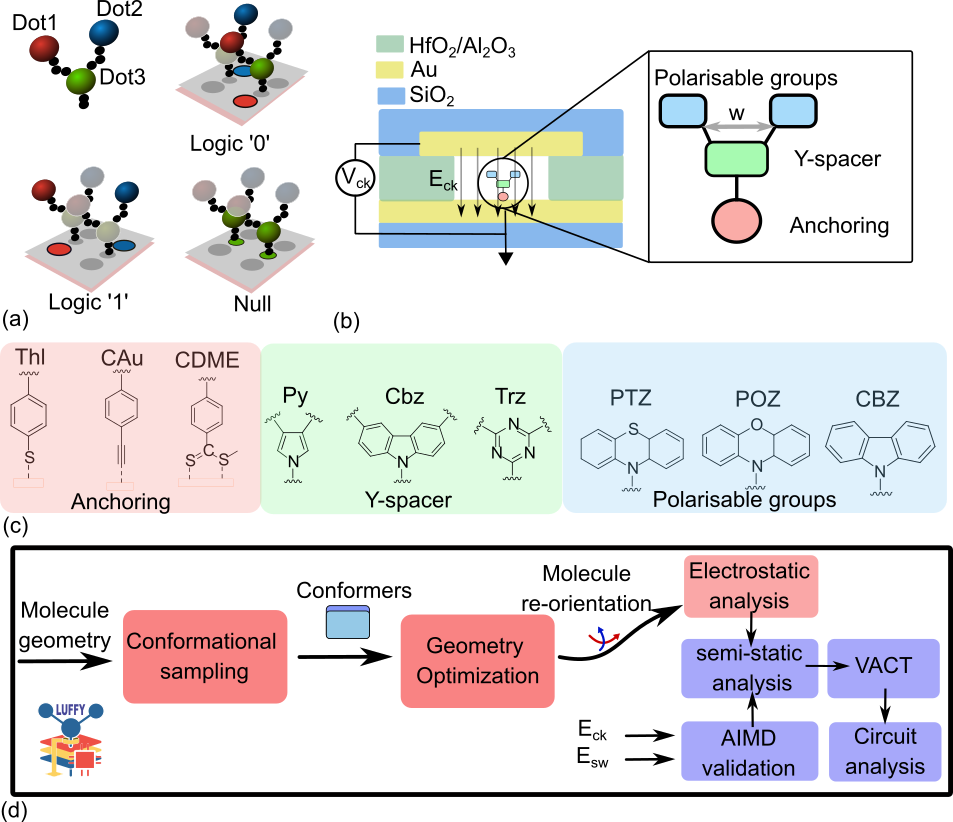}
\caption{(a) Molecular Field-Coupled Nanocomputing (MolFCN) encodes binary states through charge localization in molecular sites (“dots”), with signal propagation arising from electrostatic interactions rather than charge transport. Vertical clock fields $E_{ck}$ enable switching between the \emph{Hold} and \emph{Null} configurations. 
(b) Dielectric nano-trench architecture proposed for MolFCN implementation, where molecules are confined between a ground electrode and a closing top electrode that together define the clock field required for state switching. 
(c) Modular AG–SP–PG molecular design employed in this work: anchoring groups (AG) ensure stable and oriented surface binding, spacers (SP) enforce the Y-shaped geometry defining intramolecular dot separation, and polarizable or redox-active groups (PG) enable field-induced charge redistribution. 
(d) Overview of the unified workflow developed in this Letter, integrating conformational sampling, DFT-based electrostatic analysis, VACT derivation, dynamic validation under time-dependent fields, and SCERPA-level circuit evaluation to identify and screen viable MolFCN molecular candidates.
}
    \label{fig:Intro}
\end{figure}

%
As CMOS-technology approaches its fundamental limits in power density and switching speed, alternative computing paradigms in which information propagates without charge transport are being actively explored. Molecular Field-Coupled Nanocomputing (MolFCN) is one of these paradigms, aiming to drastically reduce energy dissipation in logic operations—as further CMOS scaling becomes increasingly constrained by thermal and switching bottlenecks \cite{anderson2014field,listo2025unfolding}. MolFCN, implementation of the Quantum-dot Cellular Automata (QCA) concept, encodes binary states through charge localization in molecular groups referred to as dots (Fig.~\ref{fig:Intro}(a)) \cite{lent2003molecular}.
MolFCN propagates logical state through purely electrostatic interactions \cite{CSLent_1993, listo2025unfolding, lu2007molecular}. Moreover, vertical fields, named clock fields $E_{ck}$, are used to enforce either the \emph{Hold} (logic-stable) or \emph{Null} (reset) configurations \cite{pulimeno2014understanding}, enabling computation with null current flow, thus ultralow power dissipation.
A concrete technological pathway towards implementation has recently been introduced via a dielectric nano-trench architecture (Fig.~\ref{fig:Intro}(b)) \cite{ravera2023roadmap, ferrero2025technology}. Although a fully functional prototype is still lacking \cite{ravera2023roadmap}, this approach represents a significant step towards bridging the gap between theoretical frameworks and experimental realization.

In this architecture, molecules are confined between a ground and a top electrode that together establish the required clock field $E_{ck}$, essential for polarizing the molecule and enabling information propagation.\cite{pulimeno2014understanding}
Operating with this platform requires molecular species with robust surface anchoring, controlled orientation, well-separated charge-localization sites, low intrinsic dipole moment, high polarizability, and chemical stability.\cite{qi2003molecular, ferrero2025technology, santana2019exploring, blair2019electric} 
Identifying and engineering molecular structures that simultaneously satisfy all these requirements remains a major open challenge, and is one of the primary reasons why a fully functional MolFCN prototype has not yet been realized \cite{ravera2023roadmap}. Several chemical families have been explored in an attempt to address these limitations, including mixed-valence complexes, Y-shaped architectures such as bis-ferrocene, and zwitterionic systems \cite{lu2005theoretical, macrae2023mixed, liza2022designing, groizard2020zwitterionic, C1NR10988J}.
Their suitability could be evaluated at the circuit level using QCADesigner \cite{walus2004qcadesigner}, which models information propagation through simplified bistable cells. 
While effective for qualitative studies, QCADesigner cannot capture the detailed molecular electrostatics that govern real molecular QCA devices \cite{ardesi2022impact, ardesi2019scerpa}.
Accurate analyses are performed using the Self-Consistent ElectRostatic Potential Algorithm (SCERPA) \cite{ardesi2019scerpa, ardesi2021scerpa}, which is embedded within the MoSQuiTo simulation framework.
MoSQuiTo explicitly incorporates molecular electrostatics by deriving Voltage-to-Aggregated-Charge Transcharacteristics (VACT) from \textit{ab initio} calculations \cite{ardesi2022impact}. In this approach, each molecule is represented by three effective point charges—aggregated charges (ACs)—that reproduce its electrostatic response under the applied clocking field $E_{ck}$ and the input potential $V_{in}$, defined as the electrostatic potential difference between the two logical dots sites \cite{ardesi2018effectiveness}.
Within this framework, SCERPA computes circuit-level information propagation by self-consistently resolving all intermolecular electrostatic interactions. This method has been validated against full DFT benchmarks \cite{ardesi2025guesstimation}.

Prior analyses have largely relied on single, energy-minimized geometries and static assumptions, neglecting conformational diversity, vibrational motion, and field-induced structural responses. However, under realistic operating conditions, molecules continuously sample a distribution of geometries due to thermal fluctuations, synthetic variability, and external fields, which can significantly affect charge localization and, consequently, logic-state stability and propagation. These factors must therefore be considered to realistically advance toward an experimental prototype.
Yet, it is well established in molecular-scale electronics that both molecular conformations and electron–vibration coupling critically affect charge localization and functional switching behavior \cite{venkataraman2006conformation, galperin2007molecular}. The absence of methodologies that account for these effects represents a key bottleneck in the realistic identification of viable MolFCN molecular candidates.
In this Letter, we propose the Layered Unified Framework for molFCN systematic analYsis (LUFFY), a VACT-based workflow that enables realistic and device-compatible molecular screening. First, \textit{conformational sampling} identifies low-energy molecular configurations, which are then refined via \textit{geometry optimization}. These optimized structures undergo \textit{single-point electrostatic analysis} to evaluate their ground-state charge distribution and polarization response. Next, VACTs are extracted using perturbation-grouping-based models. Finally, \textit{vibrational verification} assesses the molecular stability and response under different $E_{ck}$.
It then applies energy-weighted averaging to obtain conformer- and vibration-aware VACTs, and validates switching robustness through dynamic field-driven response simulations. 
This framework provides a physically grounded and scalable basis for identifying and engineering optimal MolFCN molecular candidates, thereby accelerating progress toward a functional prototype.

We apply LUFFY to a newly constructed molecular dataset specifically designed for MolFCN operation. The set includes 27 synthetically accessible Y-shaped compounds compatible with oxidation, patterning, and integration into clocked MolFCN cells. Each species of interest follows a modular architecture (Fig.~\ref{fig:Intro}(b)), where an anchoring group (AG) enables stable surface binding, a Y-spacer (SP) enforces the required dipole geometry, and a polarizable or redox-active group (PG) provides charge localization and field-driven redistribution (Fig.~\ref{fig:Intro}(c)). 
Thiols (Th), ethynyl–Au bonds (CAu), and carbodithioate methyl esters (CDME) were selected as AGs for their strong and stable gold coordination, with differing degrees of lateral mobility \cite{love2005self, herrer2019electrically, ward2024systematic}. Pyrrole (Py), carbazole (Cbz), and triazole (Trz) serve as SPs, spanning from conjugated and synthetically accessible (Py) to larger footprints (Cbz) and rigid geometries with reduced conformational freedom (Trz) \cite{shi2023recent, munir2024synthesis, yamada2018single, mishra2025multifunctional}. Carbazole (Cbz), phenoxazine (POZ), and phenothiazine (PTZ) act as PGs, offering established redox behavior and tunable polarizability \cite{munir2024synthesis, buene2021phenothiazine, luo2016recent}.
Conformational sampling in implicit water was performed using CREST code, retaining structures within 6~kcal/mol of the global minimum \cite{bannwarth2019gfn2, pracht2020automated, pracht2024crest}. 
CREST plays a crucial role in LUFFY by efficiently sampling both low- and high-energy conformers with a level of complexity beyond that of well-known conformational search methods based on classical force fields \cite{aqm,qued}. This advantage arises from its more accurate treatment of long-range interactions (electrostatics and dispersion) and its ability to incorporate solvent effects.
Indeed, CREST integrates the semi-empirical extended tight-binding method GFN2-xTB\cite{bannwarth2019gfn2} with a metadynamics-based search algorithm. 
The lowest-energy conformers for each species are reported in Fig.~S1 in the Supplementary Information (SI). The resulting conformers were optimized and analysed at the CAM-B3LYP-D4/def2-TZVP level in neutral and oxidized states using ORCA~5.0 code \cite{neese2022software, yanai2004new, WeigendDef2tzvp2005, caldeweyher2019generally}. From their charge distributions, conformer-aware semi-static $V_{in}$–$\mu$ transcharacteristics were obtained and mapped to three-point-charge representations. These were then dynamically validated under time-dependent fields \cite{ardesi2021ab} and incorporated into SCERPA \cite{ardesi2019scerpa, ardesi2021scerpa, ardesi2022impact, ardesi2025guesstimation} to evaluate information propagation under realistic intermolecular coupling. LUFFY thus provides a physically grounded route for systematic MolFCN candidate screening, bridging molecular design, electrostatic modeling, and device-level validation.

\begin{figure}[t!]
    \centering
    \includegraphics[width=0.8\linewidth]{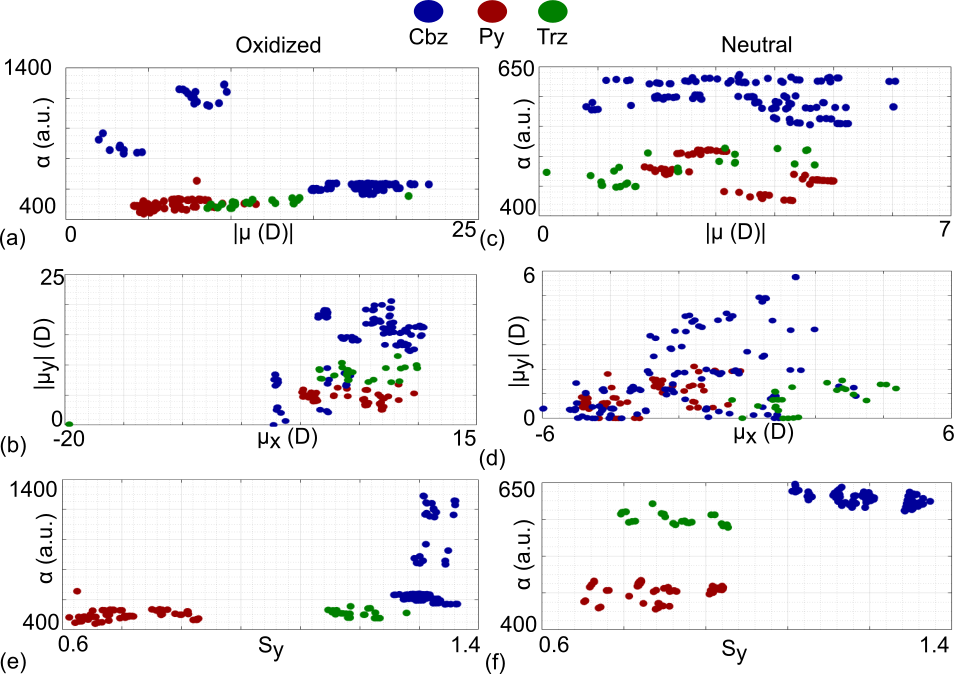}
    \caption{Conformational electrostatic and geometric descriptors for the molecular dataset. 
(a) Total dipole moment $|\mu|$ versus isotropic polarizability $|\alpha|$ (oxidized state), with each point representing a distinct conformer; Cbz-linked systems show higher $|\mu|$, $|\alpha|$, and conformational variability. 
(b) $\mu_x$ versus $|\mu_y|$ (oxidized state), where predominantly positive $\mu_x$ indicates a natural \textit{Hold} configuration; Cbz-linked compounds exhibit increased asymmetry between the two PG branches. 
(c) $|\mu|$ versus $\alpha$ in the neutral state, with Cbz remaining the most polarizable. 
(d) $\mu_x$ versus $|\mu_y|$ in the neutral state, showing a generally \textit{Null} configuration ($\mu_x<0$) and reduced dipole dispersion. 
(e) Specularity coefficient $S_y$ for the oxidized conformers, where Pyr-linked compounds show higher symmetry and Cbz-linked ones greater geometric asymmetry. 
(f) $S_y$ versus $\alpha$ for neutral conformers, displaying the same spacer-dependent clustering with shifted absolute polarizabilities. 
}
    \label{fig:SP}
\end{figure}

Understanding correlations in the high-dimensional property space of the proposed MolFCN candidates can provide valuable insights for the rational design of novel compounds \cite{butler2018machine,fromer2024algorithmic}. This strategy has already proven effective for drug-like molecules \cite{medrano23,solvaqm} and e-nose molecular building blocks \cite{moreml}.
Accordingly, we now explore the pairwise relationship between the total dipole moment $|\mu|$ and the isotropic polarizability $|\alpha|$ in the oxidized form (Fig.~\ref{fig:SP}(a)). Each point in the resulting distribution corresponds to a distinct conformer in the dataset.
 Cbz-based compounds generally display the highest $|\mu|$ and $|\alpha|$ values and the largest conformational variability, whereas Pyr yields lower values and Trz lies in between. The $\mu_x$–$|\mu_y|$ distribution in Fig.~\ref{fig:SP}(b) further shows that $\mu_x$ is predominantly positive, indicating a natural tendency toward the \emph{Hold} configuration with polarization toward the PGs. At the same time, the lateral component $\mu_y$ encodes the \emph{logical state}, as it reflects the charge imbalance between the two PG branches. Since $|\mu_y|$ is generally non-zero, most species exhibit a \emph{pre-biased logical state} at rest. The impact of this intrinsic asymmetry must be evaluated during VACT characterisation and signal propagation at the circuit level.
 Pyr-linked compounds display the lowest $|\mu_y|$, while Cbz-linked systems show a broader spread; in these compounds, the parallel increase of $\mu_x$ and $|\mu_y|$ suggests that stronger Hold polarization is accompanied by increased asymmetry between the two PG arms, leading to preferential localization on one logical branch. Grouping compounds by spacer identity clarifies these trends: the distance $w$ between the nitrogen atoms linking the PGs to the spacer increases from Pyr ($\sim$3~\AA) to Trz ($\sim$4.5~\AA) to Cbz ($\sim$8~\AA) (Fig.~S2(a)), correlating with the observed rise in $\alpha$ (Fig.~S2(b)) and the larger dispersion in both $|\mu|$ and $|\mu_y|$ (Fig.~S2(c,d)) for Cbz, reflecting greater conformational flexibility. Grouping by SP–AG–PG combinations (Fig.~S3) further shows that Cbz yields the largest number of conformers, Trz the fewest (consistent with higher rigidity), and CDME anchors systematically increase $\mu_x$ due to their stronger electron-withdrawing character (Fig.~S4). Notably, in the Cbz–Cbz family, $\mu_x$ values are systematically lower and can even become slightly negative for Th and CAu terminations, indicating a more balanced charge distribution between the two PG branches and therefore a reduced intrinsic logical bias.
Overall, the spacer primarily sets the geometric scale and polarizability, while the anchoring group biases dipole orientation. The balance between $\mu_x$ and $\mu_y$ determines whether a molecule is naturally neutral with respect to logic encoding or intrinsically favors one of the two logical states: specifically, $\mu_x$ defines the native Hold/Reset behavior, whereas $\mu_y$ sets the intrinsic logical state (0 or 1).

 For the neutral compounds, Fig.~\ref{fig:SP}(c) shows $\alpha$ in the 400–650~a.u. range with no clear correlation to $|\mu|$, though Cbz-linked species remain the most polarizable, while Fig.~\ref{fig:SP}(d) shows a weaker $\mu_x$–$|\mu_y|$ trend due to the narrower dipole distribution. Most neutrals exhibit negative $\mu_x$, indicating a natural \textit{Null} configuration with no charge polarization toward the PGs. As shown in Fig.~S5, both $w$ and $\alpha$ are largely unchanged upon oxidation, whereas dipole components reverse sign and change in magnitude. Subgroup analysis (Fig.~S6–S7) confirms that CDME anchors tend to increase $\mu_x$ also in the neutral state, and that Cbz–PTZ combinations show small oscillations around $\mu_x \approx 0$, while high-energy conformers of \textbf{M4} and \textbf{M13} display reduced $\mu_x$ in both charge states, indicating geometry-driven rather than oxidation-driven effects. To quantify how molecular symmetry is preserved across conformers, we introduce the specularity coefficient $S$ (equation \ref{eq:specularity}, details in SI), which measures deviation from mirror symmetry, where low values indicate more symmetric configurations.
 
 \begin{equation}
\begin{aligned}
S_{x,y,z} &= \frac{1}{N} \sum_{i=1}^N \left\| \vec{r}_i - \vec{r}_i^{({x,y,z})} \right\|, \\
\end{aligned}
\label{eq:specularity}
\end{equation}

As shown in Fig.~\ref{fig:SP}(e–f), Pyr-linked compounds generally exhibit more symmetry (lower $S_y$), with the two PG branches remaining nearly specular. Cbz- and, to a lesser extent, Trz-linked compounds display larger $S_y$, reflecting more pronounced Y-shaped and planar arrangements with increased conformational flexibility. Thus, $S$ provides a compact geometric descriptor that complements electrostatic analysis by enabling systematic comparison of how molecular families differ in their conformational symmetry and Y-geometry preservation.

\begin{figure}[t!]
    \centering
    \includegraphics[width=0.8\linewidth]{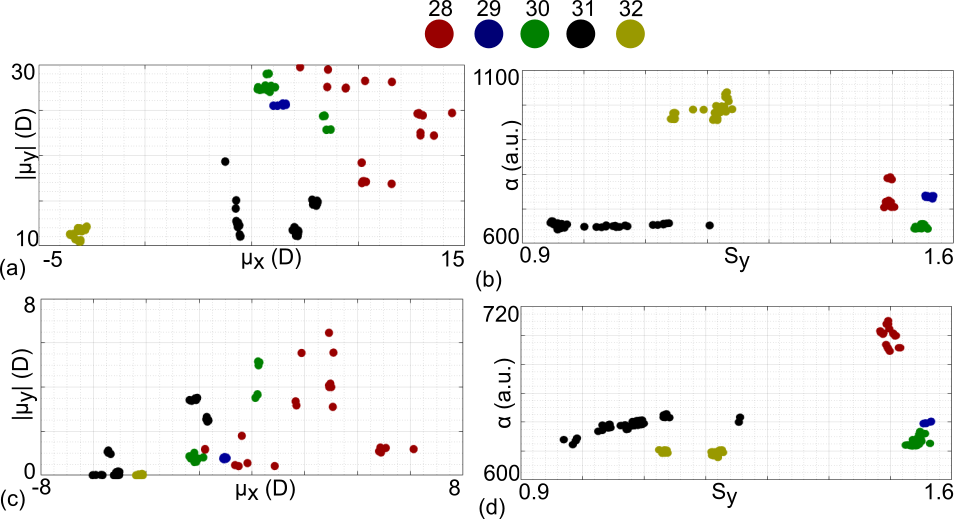}
    \caption{(a) Dipole moment component $\mu_x$ for phenyl-substituted triazole (Trz) and pyrrole (Pyr) derivatives (\textbf{M28}–\textbf{M32}) in the oxidized state. (b) Structural descriptor $S_y$ showing reduced specularity and enhanced flatness for the same set. (c) Dipole moment component $\mu_x$ for the neutral forms of molecules \textbf{M28}–\textbf{M32}. (d) $S_y$ values for neutral species, confirming the trend observed in oxidized states.}

    \label{fig:2SP}
\end{figure}

Cbz-linked compounds emerge from this analysis as the most promising candidates, combining large PG–PG separation, high polarizability, and broad conformational flexibility. Trz-linked systems, on the other hand, are more rigid, with fewer conformers and lower $\alpha$. Pyr-based species show small $w$ and $\alpha$ but remain attractive for their synthetic accessibility. To enhance Pyr and Trz derivatives, a phenyl unit—introduced for its intrinsically high $\pi$-electron polarizability—was inserted between SP and PG (compounds \textbf{M28}–\textbf{M32}, Fig.~S8). In the oxidized state, this modification increases $w$ (up to $\sim$6–8~\AA{} for Pyr and $\sim$1.2~nm for Trz, Fig.~S9(a)) and boosts $\alpha$ up to $\sim$1000~a.u. for \textbf{M32} (Fig.~S9(b)). Most derivatives show positive $\mu_x$ (Fig.~3(a)), except \textbf{M32}; Pyr-based species display low $|\mu_y|$, while Trz-linked analogues show larger fluctuations (Fig.~S9(c)). The specularity coefficient $S_y$ increases (Fig.~3(b)), indicating flatter and less mirror-symmetric Y-geometries. Notably, Trz-based molecules cluster at well-defined $S_y$ values, reflecting their intrinsically constrained rotational freedom, whereas Pyr-based derivatives exhibit a broader $S_y$ dispersion due to increased conformational flexibility. In the neutral state, $w$ remains nearly unchanged (Fig.~S10(a)), $\alpha$ increases relative to non-phenyl analogues but remains lower than in the oxidized form (Fig.~S10(b)), and $\mu_x$ spans positive to slightly negative regimes depending on the SP–PG combination (Fig.~3(c)), with \textbf{M31} and \textbf{M32} showing $\mu_y \approx 0$ except for a few polarized conformers (Fig.~S10(c)). A similar specularity trend is observed in Fig.~3(d), where Trz-linked species again group into localized $S_y$ clusters, while Pyr-linked systems display a wider spread due to additional rotational degrees of freedom. Overall, phenyl insertion provides a simple and modular strategy to enhance polarizability, tune dipole symmetry, and adjust Y-geometry, thus delivering more suitable candidates for MolFCN operation.

\begin{figure*}[t]
    \centering
    \includegraphics[width=\textwidth]{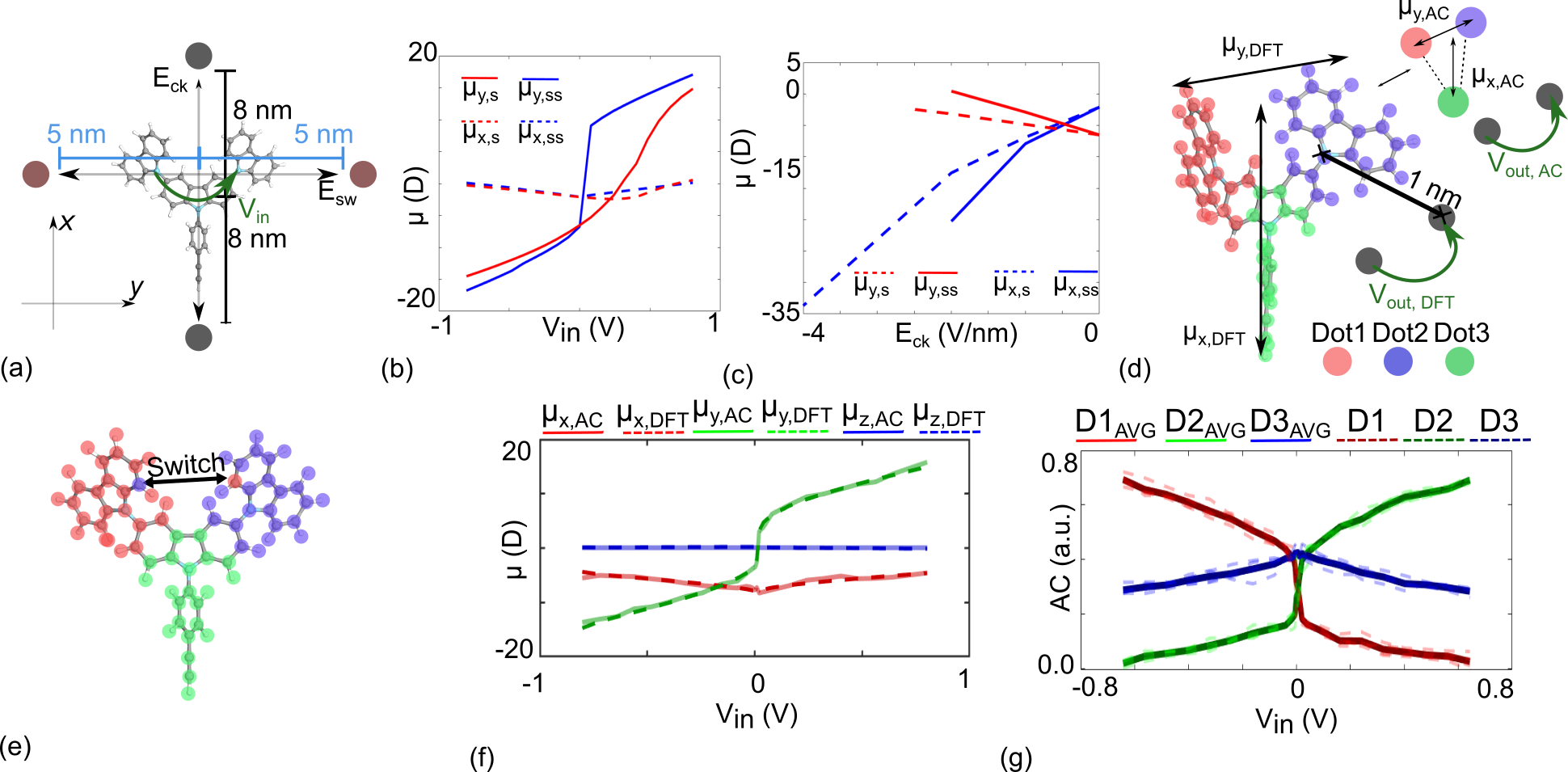}
    \caption{Analysis of \textbf{M6} with static, semi-static, and perturbation-based AC models.  
(a) Conventional static scheme for VACT extraction.  
(b--c) Comparison of $V_{in}$–$\mu_y$ characteristics: the static case ($\mu_s$) shows a smooth response, while the semi-static case ($\mu_{ss}$) yields a sharp transition near $V_{in}=0$, strongly impacting MolFCN applicability.  
(d) $E_{ck}$–$\mu_{x,y}$ curves at $V_{in}=0$, highlighting deviations of $\sim$10~D at $-2$~V/nm.  
(e) Schematic of the perturbation-based optimization, used to overcome the rigidity of conventional spatial grouping.  
(f) Validation of the perturbation-based AC model for the lowest-energy conformer, reproducing DFT $V_{in}$–$\mu$ characteristics.  
(g) Averaged VACT obtained from Boltzmann-weighted conformer contributions.  
The semi-static approach provides a more realistic description of field-induced molecular relaxation, while the perturbation-based AC and averaged VACT schemes ensure accurate reproduction of dipole components, $V_{\text{out}}$, and conformational diversity (see SI for details).}

    \label{fig:3SP}
\end{figure*}

Building on the electrostatic characterization described so far, we perform the $V_{in}$–$\mu$ analysis used for VACT extraction, as it provides the key figure of merit describing how the molecule responds to an external electric field and thereby enables its coarse-grained representation as a system of three aggregated charges. The conventional procedure assumes a fixed molecular geometry under applied $E_{ck}$ and $V_{in}$, implemented by embedding the molecule in a four-point charge environment (Fig.~4(a)), thereby neglecting field-induced nuclear rearrangements. To address this limitation, we introduce a semi-static approach: for each value of $V_{in}$, the molecular geometry is re-optimized under the applied $E_{ck}$, allowing for a structural response to the external fields. This method is applied to a representative subset of candidate species. Figs.~4(b--c) show results for \textbf{M6}, identified as the best MolFCN candidate among the analysed. The $V_{in}$–$\mu_y$ curves differ significantly: the static case ($\mu_s$) displays a smooth, weak dependence, whereas the semi-static case ($\mu_{ss}$) exhibits a sharp transition near $V_{in}=0$~V, markedly enhancing MolFCN performance. Likewise, at $V_{in}=0$~V, the $E_{ck}$–$\mu_{x,y}$ curves diverge, showing a $\sim 10$~D gap in $\mu_y$ at $E_{ck}=-2$~V/nm (Fig.~4(c)). A similar behavior is observed for \textbf{M15} (Fig.~S11(a--b)), where the rotational flexibility of the Cbz--Cbz polarizable units modulates charge transfer, making the semi-static treatment essential. By contrast, \textbf{M14}, \textbf{M9}, and \textbf{M29} (Fig.~S11(c--h)) show minimal differences between $\mu_s$ and $\mu_{ss}$, indicating limited geometry reorganization. In \textbf{M14}, $\mu_x$ remains unchanged even at $E_{ck}=-2$~V/nm (Fig.~S11(c)), as oxygen-related charge stabilization prevents induction of the \textit{Null} state. \textbf{M9} and \textbf{M29} instead achieve $\mu_y = 0$~D at $E_{ck}=-2$~V/nm, indicating proper reset behavior, similarly to \textbf{M6} and \textbf{M15}. However, \textbf{M9} does not show any $\mu_y$ sign inversion across the investigated $V_{in}$ range at $E_{ck}=0$~V/nm (Fig.~S11(f)), preventing logic-state switching and making it unsuitable for MolFCN. In contrast, \textbf{M29}, which differs only by added phenyl substituents, shows clear $\mu_y$ sign inversion between positive and negative $V_{in}$ (Fig.~S11(h)), albeit with an asymmetry reflecting a negative $V_{in}$ threshold near $-0.25$~V.
Fig.~S12 reports the neutral-state behavior of \textbf{M6} and \textbf{M29}, where positive $E_{ck}$ values are applied to evaluate \textit{Hold}-state stabilization given their intrinsically negative/low $\mu_x$. The semi-static and static treatments yield nearly identical $E_{ck}$–$\mu$ trends at $V_{in}=0$~V, with only minor deviations at $E_{ck}=2$~V/nm, and they overlap almost perfectly when varying $V_{in}$ at $E_{ck}=0$~V. Since in the neutral state the dipole variation is markedly weaker than in the oxidized configuration and insufficient to sustain robust information transfer, the following analysis focuses on oxidized molecular states, previously identified as the most effective for MolFCN information propagation~\cite{ardesi2022impact}. Moreover, considering that the semi-static approach accounts for field-induced structural rearrangements and thus provides a more physically realistic description of the molecular response, we adopt it throughout the subsequent discussion.

\begin{table*}[!t]
\centering
\caption{Semi-static dipole-component errors ($e_x,e_y,e_z$) in Debye for selected compounds in the oxidized state. Columns report, side-by-side, the first-conformer errors ($e_{x,1st,p}$, $e_{y,1st,p}$) using perturative method, the average perturbation-based grouping error ($e_{x,p},e_{y,p},e_{z,p}$), and the spatial grouping ($e_{x,sp},e_{y,sp},e_{z,sp}$).}
\label{tab:vact-validation}
\begin{tabular}{r r r r r r r r r r}
\toprule
Mol. & $E_{ck}$ & $e_{x,1st,p}$ (D) & $e_{y,1st,p}$ (D) & $e_{x,\mathrm{p}}$ & $e_{y,\mathrm{p}}$ & $e_{z,\mathrm{p}}$ & $e_{x,\mathrm{sp}}$ & $e_{y,\mathrm{sp}}$ & $e_{z,\mathrm{sp}}$ \\
\midrule
4  &  0.00  & 0.13 & 1.20 & 0.22 & 0.84 & 0.93 & 6.22 & 2.95 & 1.24 \\
29 &  0.00  & 0.19 & 0.95 & 0.15 & 0.81 & 0.20 & 4.26 & 2.72 & 0.41 \\
28 &  0.00  & 0.16 & 0.44 & 0.22 & 0.73 & 1.30 & 7.30 & 3.21 & 1.73 \\
13 &  0.00  & 0.14 & 1.38 & 0.16 & 0.86 & 0.84 & 4.59 & 2.80 & 0.89 \\
15 &  0.00  & 0.23 & 0.36 & 0.23 & 0.43 & 0.17 & 4.82 & 3.84 & 0.20 \\
15 & -1.00 & 0.14 & 0.52 & 0.18 & 0.68 & 0.36 & 0.58 & 3.59 & 0.36 \\
\hline
\hline
6  &  0.00  & 0.24 & 0.60 & 0.24 & 0.57 & 0.76 & 6.35 & 3.49 & 0.44 \\
6  & -1.00 & 0.32 & 0.35 & 0.28 & 0.48 & 0.66 & 1.63 & 2.77 & 0.58 \\
6  & -1.50 & 0.23 & 0.54 & 0.30 & 0.48 & 0.63 & 1.83 & 2.15 & 0.71 \\
6  & -2.00 & 0.59 & 0.55 & 0.73 & 0.51 & 0.57 & 7.20 & 1.36 & 0.87 \\

\bottomrule
\end{tabular}
\end{table*}

For the VACT extraction using the three-point-charge model in MoSQuiTo \cite{ardesi2018effectiveness}, validity requires that $\mu_{x,AC}, \mu_{y,AC}, \mu_{z,AC}$ and $V_{\text{out,AC}}$ reproduce the DFT-derived dipole components ($\mu_{x,DFT}, \mu_{y,DFT}, \mu_{z,DFT}$) and $V_{\text{out,DFT}}$ (Fig.~4(d)). The usual approach is to partition the molecule into three geometrically defined regions and sums the atomic charges within each one to construct the ACs (Fig.~4(d)). This spatial grouping is rigid and does not adapt to field-induced charge redistribution, thus failing to precisely capture variations under different $E_{ck}$. To overcome the limitations, we introduce a perturbation-based grouping optimization in which atomic charges are iteratively exchanged between dot1 and dot2 to reduce the $V_{\text{out,AC}}$--$V_{\text{out,DFT}}$ deviation, followed by a targeted refinement involving dot3 (Fig.~4(e); details in SI). Moreover, distinct molecular conformers may exhibit different intrinsic dipoles and different VACT responses. To account for these phenomena, we use conformer-averaged VACTs obtained via an energy-weighted combination (see sec. 3 in SI). These improvements are applied to representative candidates containing Cbz and Trz+Phenyl spacers, identified as the most polarizable and reliable linkers.

Table ~\ref{tab:vact-validation} reports the mean dipole errors along $x$, $y$, and $z$ for the analyzed species, comparing the spatial charge grouping scheme ($e_{sp}$) with the perturbation-based grouping ($e_{p}$). The reported error is defined as the average deviation between the AC-model dipoles ($\mu_{AC}$) and the DFT dipoles ($\mu_{DFT}$). For each conformer, the deviation is first averaged over the $V_{in}$ range, and the resulting values are then averaged across all conformers, according to:
\begin{equation}
\mathrm{Error}
= \frac{1}{N_{\mathrm{conf}}}
\sum_{c=1}^{N_{\mathrm{conf}}}
\left(
    \frac{1}{N_{V}}
    \sum_{i=1}^{N_{V}}
    \bigl|\mu_{AC}^{(c)}(V_i) - \mu_{DFT}^{(c)}(V_i)\bigr|
\right),
\end{equation}
where $N_{\mathrm{conf}}$ is the number of conformers, $N_{V}$ is the number of sampled $V_{in}$ values, and $\mu_{AC}^{(c)}(V_i)$ and $\mu_{DFT}^{(c)}(V_i)$ are the AC-model and DFT dipoles, respectively, computed for conformer $c$ under the input potential $V_i$.
In the spatial grouping, atomic charges are assigned to three dots by partitioning the molecular $y$-coordinates with a fixed 2~\AA\ threshold around the molecular center: dot1 and dot2 correspond to the nitrogen atoms of the polarizable groups, while dot3 is placed on the carbon atom of the anchoring group bound to the SP. The perturbation-based grouping uses the spatial scheme only as an initial guess and then automatically refines the grouping through charge exchanges, consistently yielding smaller errors. As a result, $e_{p}$ is significantly lower than $e_{sp}$ in all cases (Table~\ref{tab:vact-validation}). The table also reports the $x$- and $y$-dipole errors for the lowest-energy conformer, which are below 1~D for $\mu_x$ and around 1~D for $\mu_y$ in all cases, confirming high accuracy. While spatial grouping could in principle be manually fine-tuned to reduce these errors, doing so requires manual redefinition of the initial charge groupings, making the approach time-consuming and not robust; moreover, because the grouping is fixed by construction, it lacks flexibility under varying field conditions ($E_{ck}$,$V_{in}$) and can therefore remain inaccurate in capturing field-induced charge redistribution. In contrast, the perturbation-based scheme performs this adjustment automatically and dynamically, allowing the charge grouping to adapt to the electrostatic environment, which results in consistently higher accuracy without manual intervention. Fig.~4(f) illustrates the perturbation-based accuracy for the first conformer of \textbf{M6} at $E_{ck}=-1$~V/nm, where the $V_{in}$–$\mu$ transcharacteristics match the DFT reference.

In Fig.~S13(a), the individual conformer $V_{in}$–$\mu$ curves are shown as light solid lines, while the energy-weighted average—almost overlapping due to minimal conformer energy differences (weights: 0.251, 0.250, 0.249, 0.249)—is shown as dashed lines. This result confirms that a set of three aggregated charges can faithfully reproduce the electrostatic response, enabling automatic and reliable VACT extraction, as reported in Fig.~4(g).
Figs.~S13(b)--(d) details, respectively, the first-conformer analysis, the extracted VACT, and the DFT-extracted conformer-averaged $V_{in}$–$\mu$ characteristics at $E_{ck}=0$, $-1.5$, and $-2$~V/nm. At $E_{ck}=0$~V/nm, the VACT closely resembles that at $-1$~V/nm but displays a slightly stronger asymmetry around $V_{in}=0$~V, arising from conformers with opposite-sign native $\mu_y$ that average toward intermediate values. This observation highlights how conformational diversity influences the effective MolFCN response. At stronger fields, charge redistribution becomes dominant: starting from $E_{ck}=-1.5$~V/nm, dot3 gradually accumulates positive charge, and by $E_{ck}=-2$~V/nm the system is effectively in the reset condition, with $\mu_y$ nearly suppressed and most positive charge localized on dot3, while the other dots contribute negligibly. This behavior demonstrates to a robust \textit{Null} state, essential for reliable MolFCN operation.
Fig.~S14 reports the $E_{ck}=0$~V/nm results for \textbf{M15} (a), \textbf{M13} (b), \textbf{M4} (c), \textbf{M28} (d), and \textbf{M29} (e). The lowest-energy conformers exhibit small $e_{x,1st,\mathrm{p}}$ and $e_{y,1st,\mathrm{p}}$ values (Tab.~\ref{tab:vact-validation}), confirming the accuracy of the perturbation-based grouping. \textbf{M15}, \textbf{M28}, and \textbf{M29} show asymmetric $V_{in}$--$\mu$ curves around $V_{in}=0$~V due to a native dipole, yet the dipole reverses at modest biases ($|V_{in}|\!<\!0.5$~V), indicating accessible switching. In contrast, \textbf{M4} and \textbf{M13} (with opposite-sign responses) require higher inputs ($|V_{in}|\!\sim\!0.6$~V), consistent with a more stable native logic state, likely influenced by the higher electronegativity of the PTZ unit. \textbf{M28} remains more easily switchable, plausibly due to phenyl-induced charge delocalization. Fig.~S15 shows \textbf{M15} under $E_{ck}=-1$~V/nm, where the perturbation-based grouping remains valid (Tab.~\ref{tab:vact-validation}). The semi-static energy-averaged VACT and the corresponding $V_{in}$--$\mu$ curve fall between those of the individual conformers, reflecting the near-symmetric switching of the conformer ensemble near the transition. As in \textbf{M6}, the applied field promotes partial charge localization on dot3, yielding an intermediate operating point between \textit{Hold} and \textit{Null}.

Across multiple cases, the energy-averaged characteristics lies between those of the single conformers, directly reflecting the influence of conformational diversity: different geometries favor different native charge states, and their ensemble naturally leads to intermediate responses. This observation indeed raises a central modeling question regarding whether VACT should be taken from the single lower-energy conformer or from the conformer group, since accurate information propagation analysis depends critically on the chosen VACT.
As a practical criterion, we compare the MD-derived steady-state dipole $\langle \Delta\mu \rangle_t$ to the predictions of both the lowest-energy conformer VACT and the ensemble-averaged VACT. Specifically, we evaluate the deviation of $\langle \Delta\mu \rangle_t$ from the corresponding $V_{in}$--$\mu$ curves. A smaller deviation for the ensemble-averaged VACT indicates that multiple conformers are being sampled under the applied $(V_{in},E_{ck})$ conditions, and therefore the ensemble-averaged VACT provides a physically relevant description. In contrast, if the lowest-energy conformer yields the smaller deviation, the molecule remains confined near its initial ground-state geometry and the single-conformer VACT is appropriate.

\begin{figure*}[t]
    \centering
    \includegraphics[width=\textwidth]{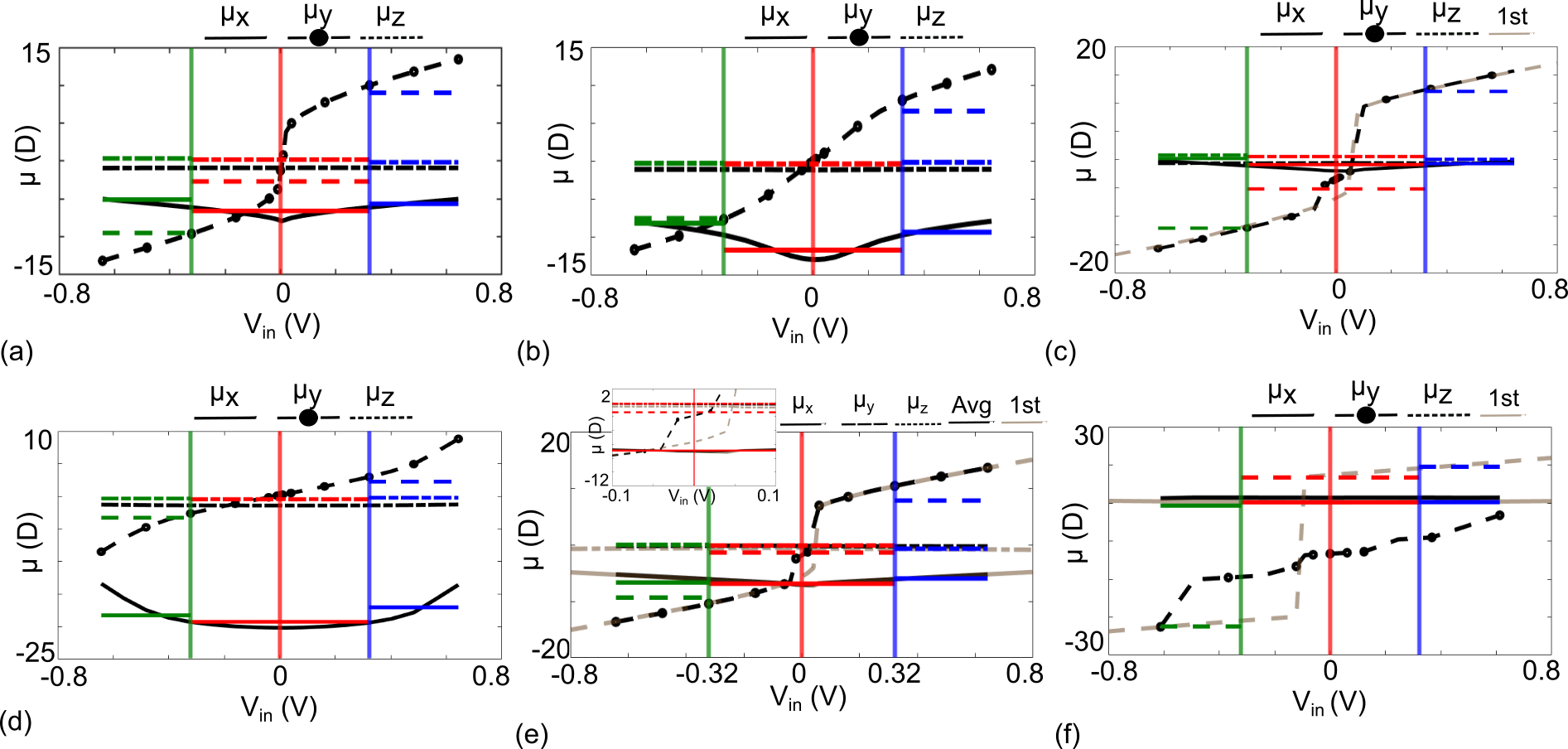}
    \caption{(a) $V_{in}$–$\mu$ transcharacteristics of \textbf{M6} at $E_{ck}=-1$ V/nm; dashed vertical lines mark the AIMD bias points ($V_{in}=-0.32 V,0,+0.32 V$), with colored horizontal lines indicating time–averaged dipoles and black curves the energy–averaged VACT. (b) $E_{ck}=0$ V/nm case for \textbf{M6}, highlighting deviations in $\mu_y$ between AIMD averages and energy–averaged curves. (c) $E_{ck}=-2$ V/nm case, where AIMD confirms the stabilization of a robust \textit{Null} state. (d) $V_{in}$–$\mu$ curves of \textbf{M28} at $E_{ck}=0$ V/nm, showing closer agreement of AIMD results with the lowest–energy conformer rather than the averaged characteristic. (e) SCERPA propagation for \textbf{M6} using averaged VACTs at –1, –1.5, and –2 V/nm: snapshots of logic inputs ‘0’ and ‘1’ reaching the output in a three–zone, clocked molecular circuit (see Fig. S17 for full dynamics).}
    \label{fig:3SP}
\end{figure*}

\begin{table}[t]
\centering
\caption{Absolute errors $\epsilon = |\mu^{\text{AIMD}} - \mu^{\text{ref}}|$ (Debye) between time-averaged AIMD dipoles ($\mu^{\text{AIMD}}$) and reference values from the $V_{in}$–$\mu$ curves. Two sets of reference values are considered: the energy-averaged curve ($\epsilon_{avg}$) and the first conformer ($\epsilon_{1st}$).}
\begin{tabular}{ccccccccc}
\toprule
\multirow{2}{*}{Mol.} & \multirow{2}{*}{$E_{ck}$ [V/nm]} & \multirow{2}{*}{$V_{in}$ [V]} & \multicolumn{3}{c}{$\epsilon_{avg}$ [D]} & \multicolumn{3}{c}{$\epsilon_{1st}$ [D]} \\
\cmidrule(lr){4-6}\cmidrule(lr){7-9}
 & & & $x$ & $y$ & $z$ & $x$ & $y$ & $z$ \\
\midrule
\multirow{12}{*}{6} 
 & \multirow{3}{*}{-1} 
   & -0.32 & -0.16 & 0.11 & 1.25 & 1.20 & 0.33 & 0.19 \\
 &  &  0.00 & 1.28 & 0.74 & 1.01 & 1.34 & 0.92 & 0.06 \\
 &  & +0.32 & 0.51 & 0.80 & 0.61 & 0.59 & 0.81 & 0.29 \\
\cmidrule(lr){2-9}
 & \multirow{3}{*}{-1.5}
   & -0.32 & 1.54 & 0.13 & 0.81 & 1.70 & 0.26 & 0.25 \\
 &  &  0.00 & 1.24 & 0.29 & 0.76 & 1.32 & 0.40 & 0.38 \\
 &  & +0.32 & 0.38 & 1.43 & 0.92 & 0.51 & 1.43 & 0.03 \\
\cmidrule(lr){2-9}
 & \multirow{3}{*}{0}
   & -0.32 & 1.29 & 0.02 & 1.41 & 1.39 & 0.42 & 0.29 \\
 &  &  0.00 & 1.11 & 1.73 & 1.09 & 1.24 & 1.63 & 0.02 \\
 &  & +0.32 & 0.40 & 0.30 & 0.63 & 0.47 & 0.25 & 0.28 \\
\cmidrule(lr){2-9}
 & \multirow{3}{*}{-2}
   & -0.32 & 1.04 & 0.66 & 1.05 & 1.11 & 0.95 & 0.28 \\
 &  &  0.00 & 0.91 & 0.64  & 0.98 & 1.02 & 0.89 & 0.42 \\
 &  & +0.32 & 2.35 & 0.72 & 1.17 & 2.56 & 0.90 & 0.21 \\
 \midrule
\multirow{3}{*}{15}
 & \multirow{3}{*}{-1}
   & -0.32 & 0.57 & 1.09 & 0.172 & 0.52 & 1.01 & 0.58 \\
 &  &  0.00 & 0.15 & 0.487 & 0.067 & 0.174 & 4.28 & 0.41 \\
 &  & +0.32 & 0.12 & 2.57 & 0.43 & 0.12 & 2.55 & 0.04 \\
\midrule
\multirow{3}{*}{28}
 & \multirow{3}{*}{0}
   & -0.32 & 2.12 & 13.05 & 1.43 & 0.98 & 1.49 & 1.12 \\
 &  &  0.00 & 1.18 & 20.05 & 0.06 & 0.04 & 0.66 & 0.36 \\
 &  & +0.32 & 1.17 & 18.80 & 0.44 & 0.01 & 0.43 & 0.56 \\
\bottomrule
\end{tabular}
\label{tab:errorsAll}
\end{table}

Fig.~5(a) shows the $V_\text{in}$–$\mu$ characteristics of \textbf{M6} under $E_{ck}=-1$~V/nm. Vertical dashed lines indicate the bias values used in AIMD ($V_{in}=-0.32$~V, $0$~V, and $+0.32$~V); horizontal colored lines mark the corresponding time-averaged dipoles, while black curves denote the ensemble-averaged VACT. Solid, dotted, and densely dashed traces correspond to $\mu_x$, $\mu_y$, and $\mu_z$, respectively. The deviations between the AIMD averages and the ensemble-averaged VACT are quantified as $\varepsilon_{\text{avg}}$, while those relative to the lowest-energy conformer VACT are denoted $\varepsilon_{\text{1st}}$, and are reported in Table~\ref{tab:errorsAll}. Fig.~S16 shows the time evolution of the dipole for the three applied biases.
For $E_{ck}=-1$~V/nm (Fig.~5(a)), the discrepancies at the three bias points are similar for both the energy-averaged ($\epsilon_{\mathrm{avg}}$) and lowest-energy conformer ($\epsilon_{\mathrm{1st}}$) predictions, with the ensemble-averaged curve overall aligning well with the AIMD time averages. The dipole dynamics show that at $V_{in}=+0.32$~V the dipole reaches steady state only after $\sim$100~fs (Fig.~S16(c)), unlike the faster stabilization at $V_{in}=0$ and $-0.32$~V (Fig.~S16(a),(b)), because the native $\mu_y$ initially opposes the field-induced dipole and the geometry must reorganize before settling. A similar delayed settling is observed at $E_{ck}=-1.5$~V/nm (Fig.~5(b)), where the corresponding $\mu(t)$ traces (Fig.~S16(d)–(f)) again show transient relaxation at $V_{in}=+0.32$~V. These transient-response behaviors directly inform time-domain MolFCN modeling directly providing the limits of the intrinsic switching speed \cite{listo2024unveiling, ardesi2021ab}. For $E_{ck}=0$~V/nm (Fig.~5(c)), a larger deviation appears in $\mu_y$ at $V_{in}=0$~V, with $\epsilon_{y,\mathrm{avg}}$ and $\epsilon_{y,\mathrm{1st}}$ nearly identical and both exceeding 1~D; at $V_{in}=\pm0.32$~V, the errors from both references become comparable as the system enters a saturation regime where the semi-static $V_{in}$–$\mu$ curves converge toward the AIMD averages (Fig.~S16(g)–(i)). For $E_{ck}=-2$~V/nm (Fig.~5(d)), $\epsilon_{\mathrm{avg}}$ and $\epsilon_{\mathrm{1st}}$ at $V_{in}=0$~V remain nearly identical, while the residual deviation in $\mu_x$ originates from the $\sim$150~fs relaxation needed to reach $\approx -20$~D (Fig.~S16(l)); at such values of electric field, $\mu_y$ is largely quenched while $\mu_x$ stabilizes a robust \textit{Null} state. A similar settling behavior occurs at $V_{in}=\pm0.32$~V (Fig.~S16(m),(n)), where the dipoles converge close to the semi-static prediction, with residual errors of $\epsilon_{x,\mathrm{avg}}=1.04$, $\epsilon_{y,\mathrm{avg}}=0.66$ at $V_{in}=-0.32$~V, and $\epsilon_{x,\mathrm{avg}}=2.35$, $\epsilon_{y,\mathrm{avg}}=0.72$ at $V_{in}=+0.32$~V, reinforcing that finite settling times must be accounted for when estimating clock frequency. Comparable behavior is observed for \textbf{M15} at $E_{ck}=-1$~V/nm (Fig.~5(e), Fig.~S17(a)–(c)): at $V_{in}=0$~V, AIMD reveals large $\mu_y$ oscillations while $\mu_x$ remains stable, and at $V_{in}=\pm0.32$~V both dipole components agree well with the ensemble-averaged semi-static model. Notably, here the ensemble-averaged $V_{in}$–$\mu$ reproduces the AIMD time average more accurately than the first-conformer curve, indicating that incorporating conformational sampling leads to a more accurate semi-static model.

\begin{figure}[t]
    \centering
    \includegraphics[width=0.7\linewidth]{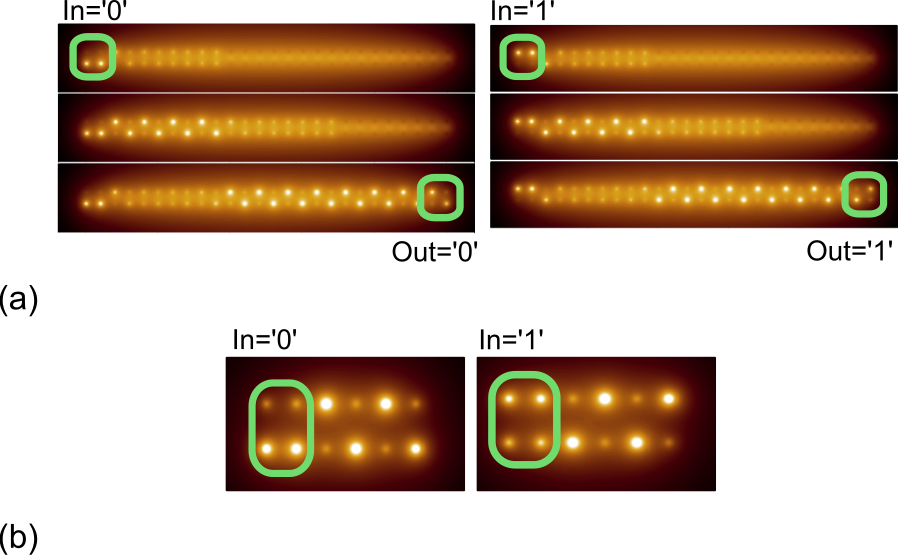}
    \caption{(a) Propagation of logical inputs ‘0’ and ‘1’ through three 8–molecule clocking zones (1 nm spacing) driven by the signals in Fig. S19(a), confirming the suitability of \textbf{M6} for MolFCN operation. (b) Minimal six-cell wire based on \textbf{M28} with reduced spacing (0.8 nm) to counteract VACT asymmetry, showing correct propagation under $\pm 1.3$ V drivers and demonstrating its potential as a MolFCN candidate.}
    \label{fig:prop}
\end{figure}

The opposite trend is observed for \textbf{M28} (Fig.~5(f)): at $E_{ck}=0$~V/nm, the energy-averaged $\mu_{x,y}$ curves (black) deviate from the AIMD time-averaged dipoles at $V_{in}=0$, $-0.32$, and $+0.32$~V (see the corresponding $\mu(t)$ traces in Fig.~S18(a)–(c)). The $V_{in}$–$\mu$ characteristics of the lowest-energy conformer (light brown in Fig.~5(f)) closely reproduce the AIMD averages, as also reflected by the lower $\epsilon_{\mathrm{1st}}$ values reported in Table~\ref{tab:errorsAll}.
Taken together, these results show that the agreement between AIMD time-averaged dipoles and the semi-static model (either single-conformer or ensemble-averaged) provides a reliable criterion for selecting the appropriate VACT representation. The transient dipole dynamics thereby validate the semi-static information–propagation framework within the relevant stabilization timescales. Future work will extend this approach to substrate-aware modelling, capturing charge redistribution and atomic relaxation of anchored molecules to enhance device-level realism \cite{koch2013molecule}.

To complete the analysis, we use the AIMD-validated averaged VACTs for \textbf{M6} at $E_{ck}=-1$, $-1.5$, and $-2$~V/nm as inputs to SCERPA for propagation studies, thereby assessing its suitability for MolFCNs. Fig.~6(a) shows three propagation steps of logical ‘0’ and ‘1’, which are stably transmitted to the output; full propagation traces are reported in Fig.~S19. The wire consists of three clocking zones of eight molecules each, with a spacing of 1~nm and driven by the signals in Fig.~S19(a). This result closes the loop on the first systematic validation workflow of a MolFCN molecular candidate—from conformational and electrostatic characterization, to semi-static modelling, to dynamical validation, and finally to circuit-level propagation—demonstrating strong potential for \textbf{M6} in MolFCN architectures. Fig.~6(b) presents a minimal wire architecture based on \textbf{M28}, composed of six cells. Due to the intrinsic VACT asymmetry (Fig.~S14(b), Fig.~5(f)), which biases one logic state, the intermolecular spacing was reduced to $0.8$~nm to enhance electrostatic coupling. Using the lowest-energy conformer VACT validated at $E_{ck}=0$~V/nm and applying $\pm 1.3$~V to the drivers, we observe a desirable propagation of logical ‘0’ and ‘1’. This result constitutes a first, preliminary demonstration of information transport using \textbf{M28}, showing that, despite its native VACT asymmetry, it remains a promising candidate for MolFCN implementations.

Overall, our work provides the first framework for the rational design and validation of molecular candidates for MolFCN architectures, bridging quantum-informed molecular properties with functional device behavior. By explicitly accounting for conformational variability and its effect on electrostatic response, LUFFY enables the consistent and accurate identification of molecules capable of supporting stable information transport.
Applied to a dataset of 27 compounds, the framework identifies carbazole-based Y-spacers as the most promising class and, crucially, reveals two molecular candidates whose electrostatic transcharacteristics remain robust across conformations. When introduced into SCERPA, these candidates enable reliable information propagation. These findings demonstrate that conformationally averaged electrostatic descriptors are essential for predicting functionality in MolFCN systems. In addition, the systematic advantage of the oxidized state further refines the molecular design space.
Beyond candidate identification, LUFFY establishes a transferable strategy for connecting molecular design with circuit-level functionality. Future developments will incorporate substrate-induced effects through substrate-aware VACTs, enabling more realistic modelling of molecule–environment interactions. The framework is also inherently compatible with sustainable machine learning practices \cite{susml} and can be extended to automate the extraction of optimal transcharacteristics and accelerate high-throughput screening. Together, these advances open a pathway toward scalable, AI-assisted discovery of MolFCN building blocks.

\section*{Acknowledgments}
Funded by the Deutsche Forschungsgemeinschaft (DFG, German Research Foundation) under Germany's Excellence Strategy - EXC 3115 - 533767731.
Funding by the DFG via the “Responsible Electronics in the Climate Change Era – REC²” Cluster of Excellence (EXC 3035, Project-ID 533607596) is gratefully acknowledged.
Funded by the German Research Foundation (DFG, Deutsche Forschungsgemeinschaft) as part of Germany’s Excellence Strategy – EXC 2050/2 – Project ID 390696704 – Cluster of Excellence “Centre for Tactile Internet with Human-in-the-Loop” (CeTI) of Technische Universität Dresden.
We thank the Center for Information Services and High-Performance Computing (ZIH) at TU Dresden for providing the computational resources and technical support.


\section*{Supporting Information}

Additional calculations and analyses supporting the conclusions of this work are provided in the online Supplementary Information.



\bibliography{molfcn}

\section*{Data availability}
The dataset generated in this work and scripts to use the LUFFY workflow are available in the LUFFY GitHub repository (https://github.com/federicoravera-sudo/LUFFY).

\section*{Author contributions}
The work was initially conceived by FR and LMS, with additional design contributions from AV and YA.
FR and LMS developed the QM dataset and defined the structure of the LUFFY framework.
AV proposed the molecular building blocks used to generate the dataset.
FR performed the molecular simulations using the LUFFY framework, analyzed the results, and developed the GitHub repository under the supervision of LMS and YA.
All authors discussed the results and contributed to the final manuscript.

\end{document}


\maketitle

\clearpage

\section*{1. Molecular Dataset}
\begin{figure*}[h]
    \centering
    \includegraphics[width=\textwidth]{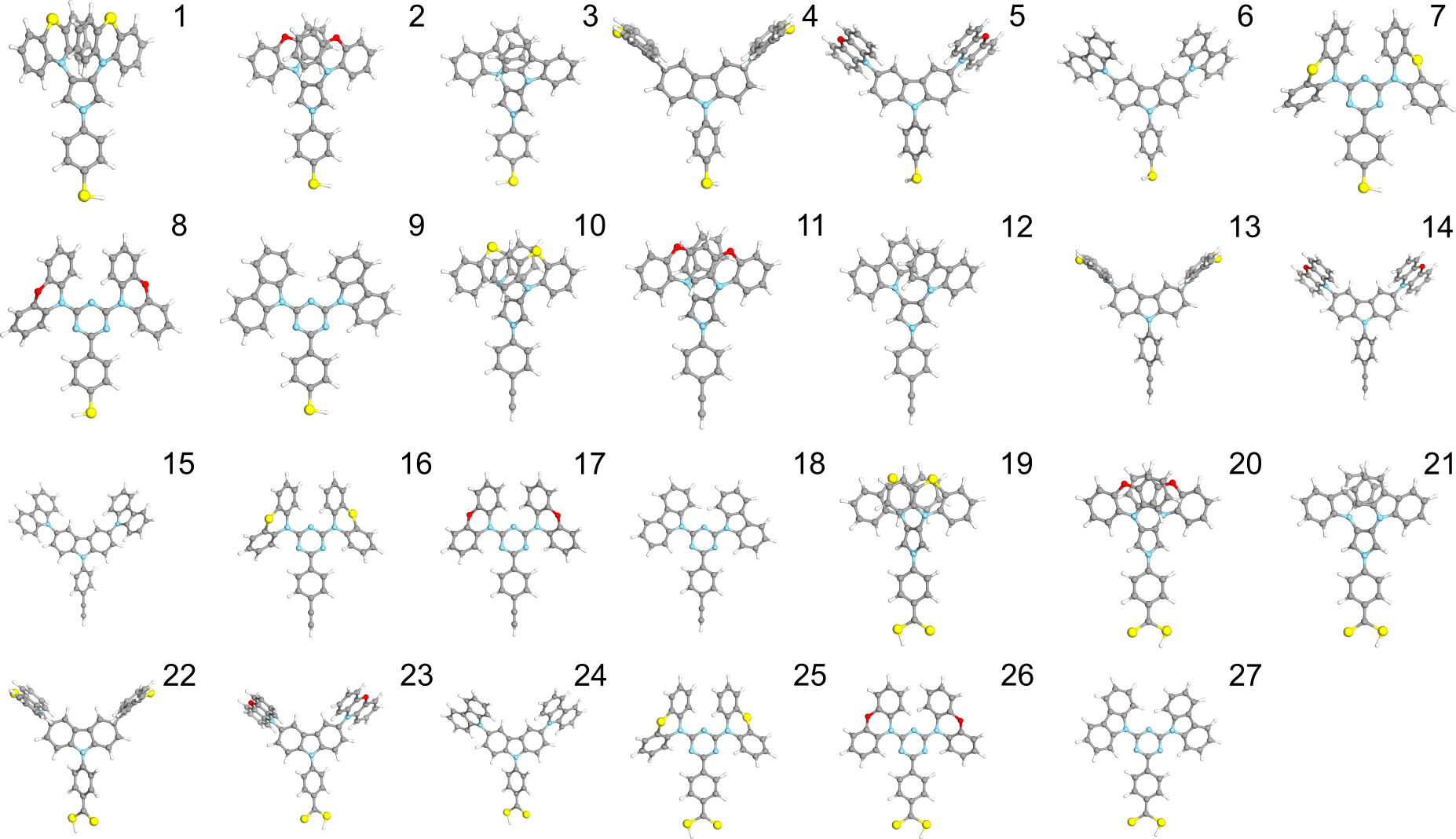}
    \caption{Molecular dataset presentation. Geometry of the best conformer for the proposed molecules.}
    \label{fig:SI1}
\end{figure*}

\clearpage

\section*{2. Oxidized-State Electrostatic Properties}
\begin{figure}[h]
\centering
\includegraphics[width=\columnwidth]{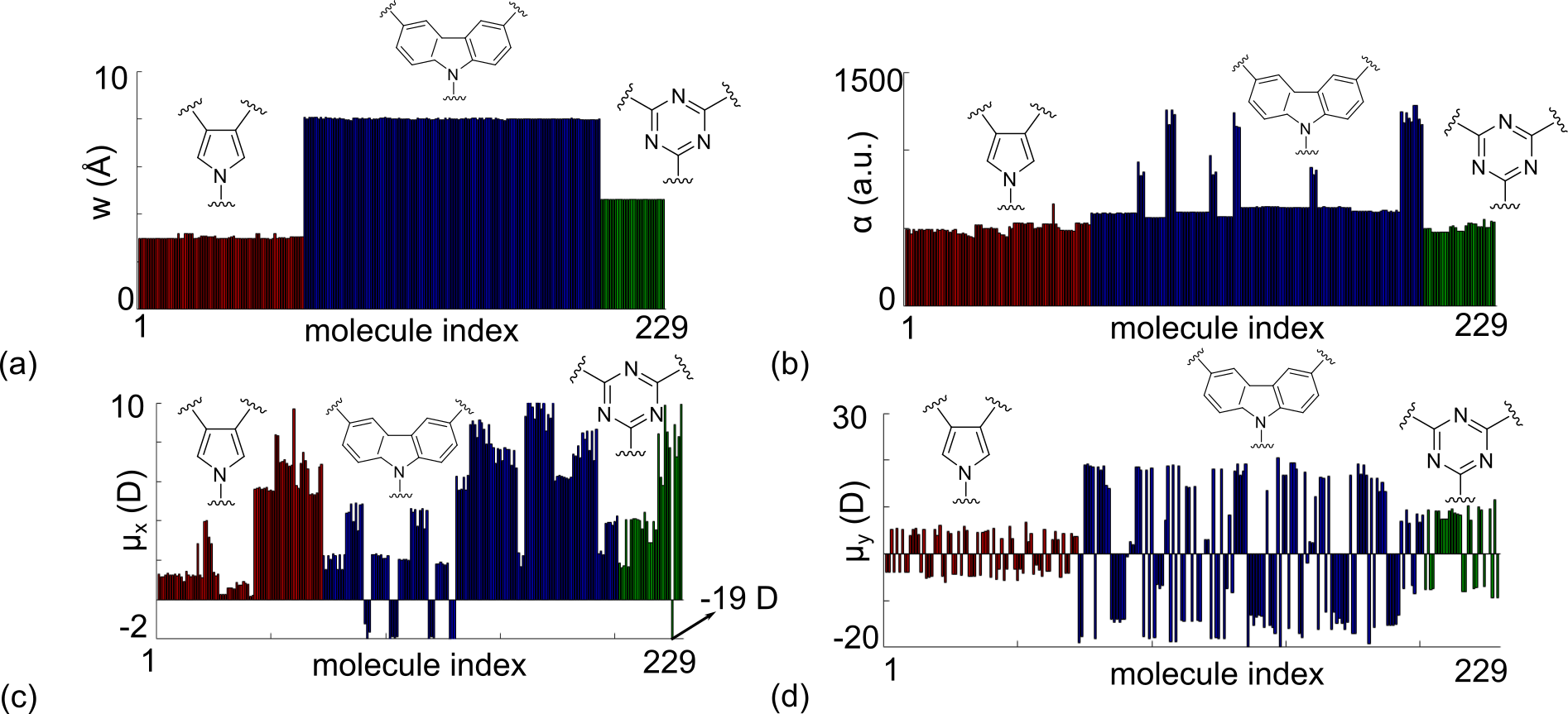}
\caption{Electrostatic descriptors for oxidized conformers.}
\label{fig:SI2}
\end{figure}

\begin{figure}[h]
\centering
\includegraphics[width=\columnwidth]{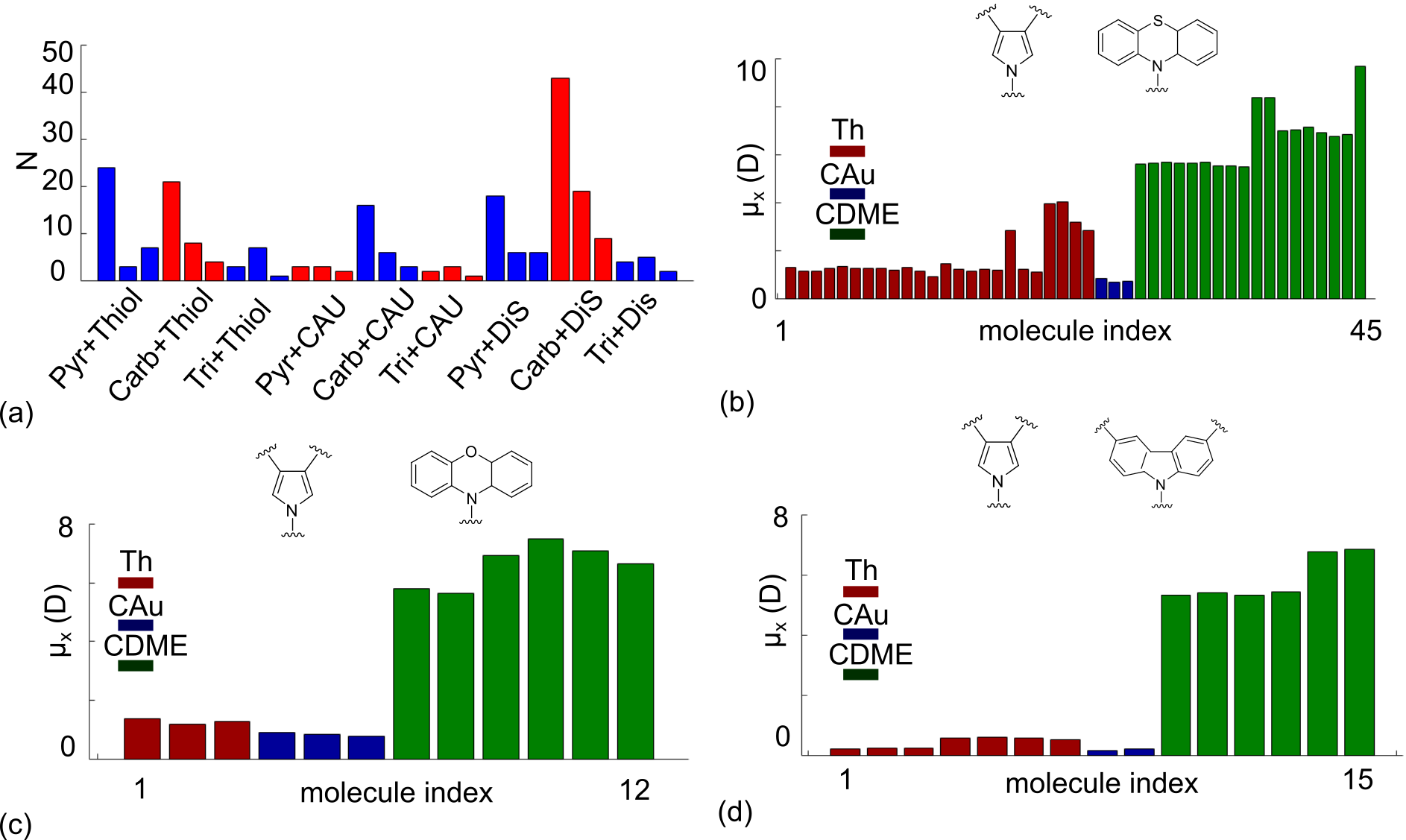}
\caption{Conformer count and $\mu_x$ trends for different anchor–spacer combinations.}
\label{fig:SI3}
\end{figure}

\begin{figure}[h]
\centering
\includegraphics[width=\columnwidth]{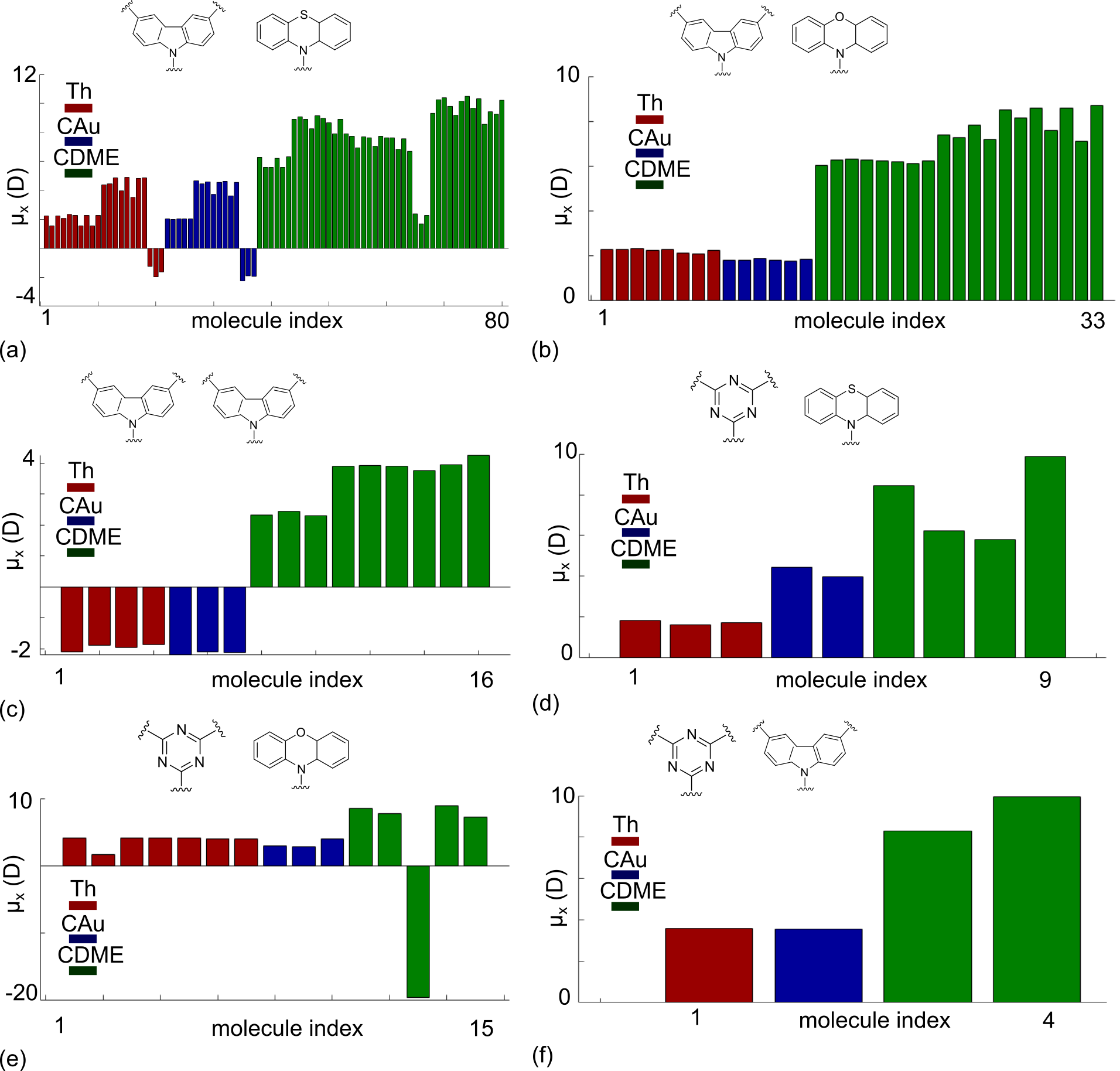}
\caption{Oxidized-state $\mu_x$ trends as a function of anchoring group (AG).}
\label{fig:SI4}
\end{figure}

\clearpage

\section*{3. Neutral-State Electrostatic Properties}
\begin{figure}[h]
\centering
\includegraphics[width=\columnwidth]{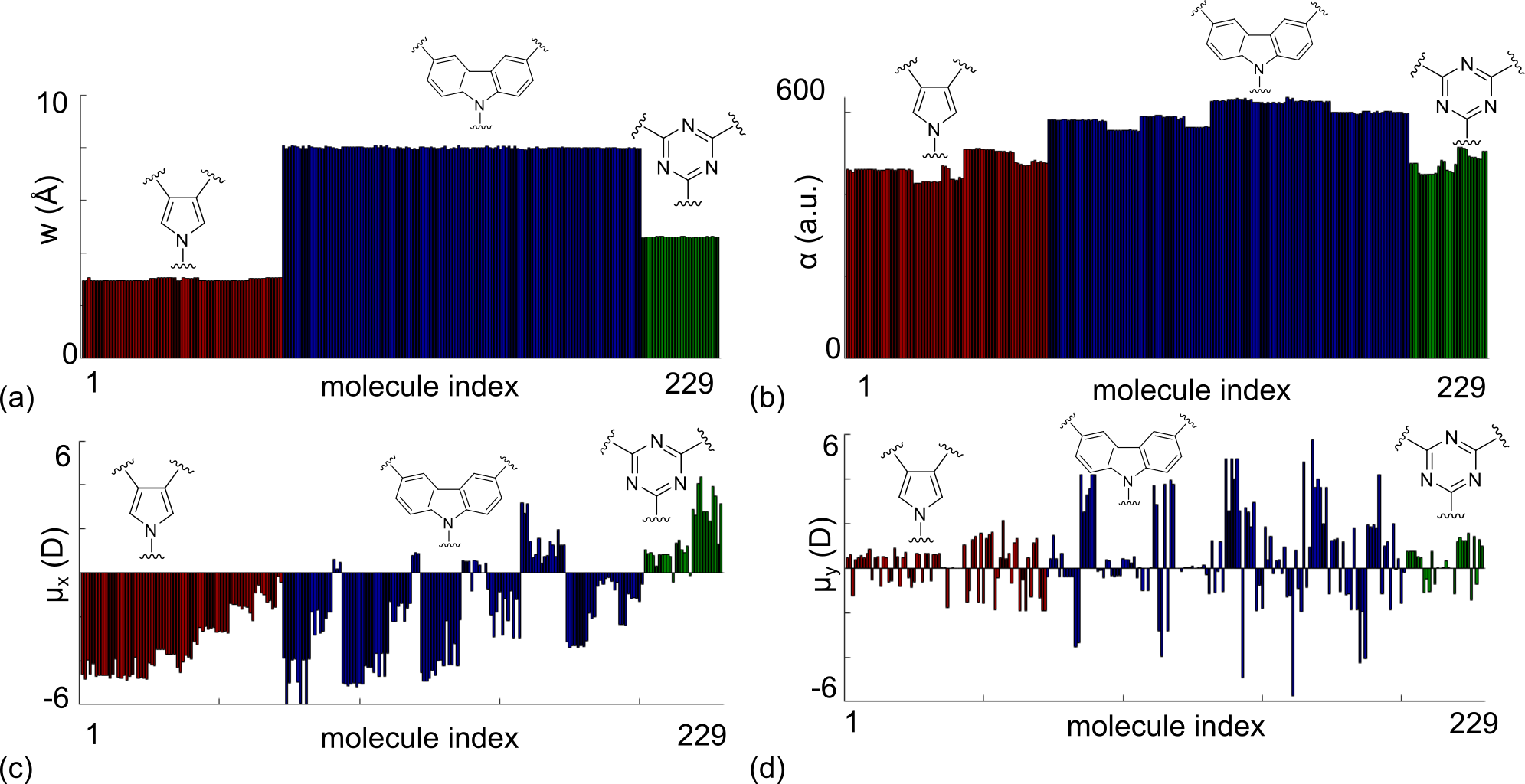}
\caption{Neutral molecules electrostatic analysis.}
\label{fig:SI5}
\end{figure}

\begin{figure}[h]
\centering
\includegraphics[width=\columnwidth]{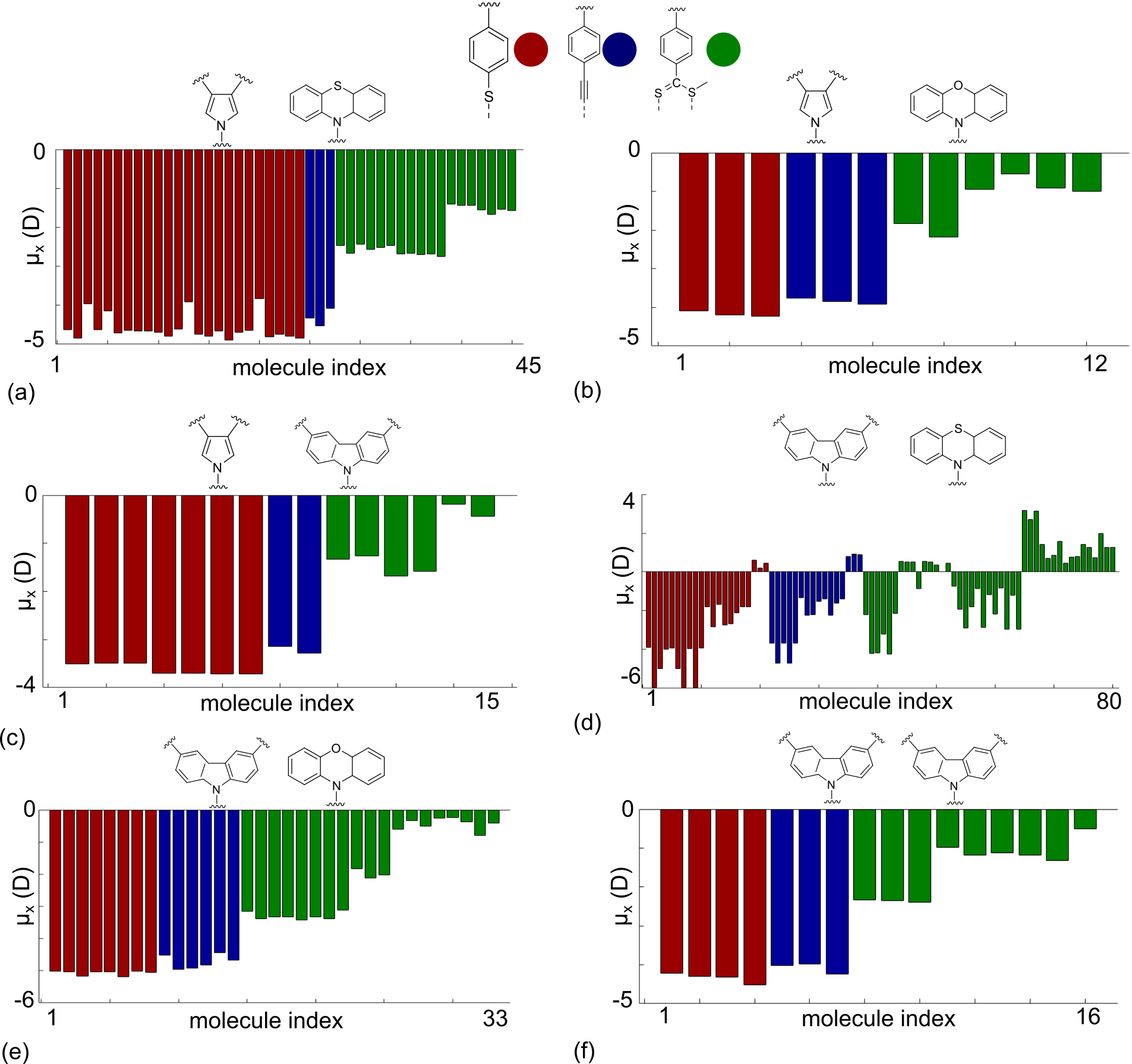}
\caption{Neutral molecules: $\mu_x$ variations for selected PG–SP combinations.}
\label{fig:SI6}
\end{figure}

\begin{figure}[h]
\centering
\includegraphics[width=\columnwidth]{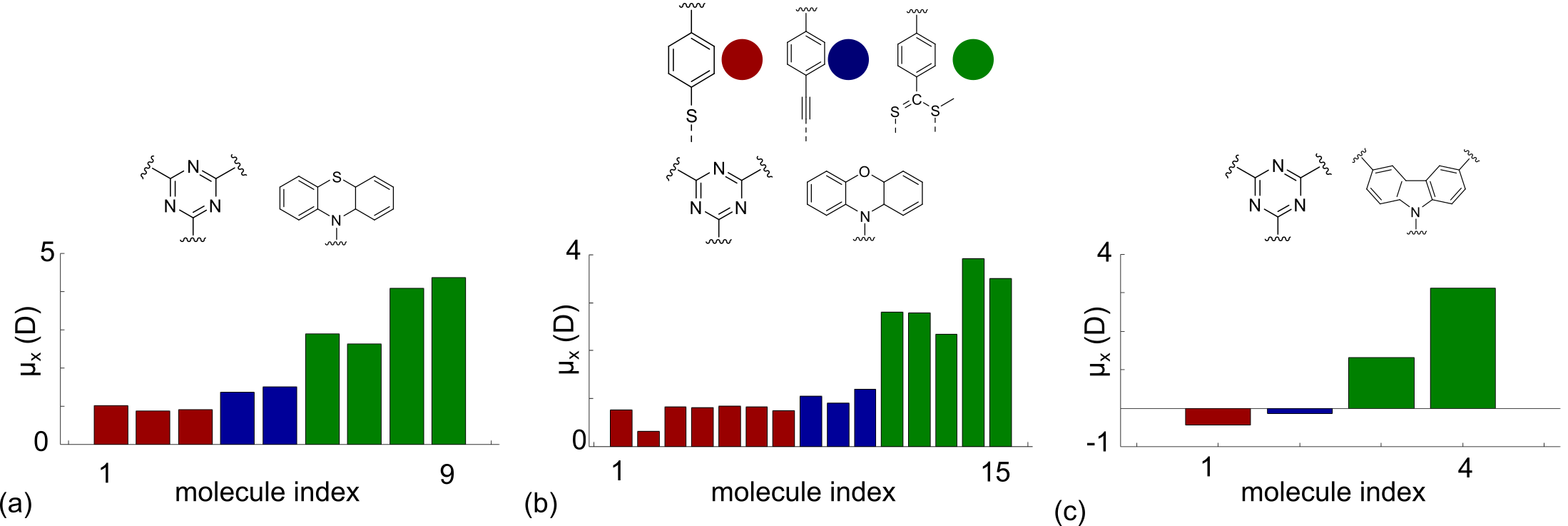}
\caption{Neutral molecules: $\mu_x$ variation for Trz-linked derivatives.}
\label{fig:SI7}
\end{figure}

\clearpage

\section*{4. Phenyl-Substituted Derivatives}
\begin{figure}[h]
\centering
\includegraphics[width=\columnwidth]{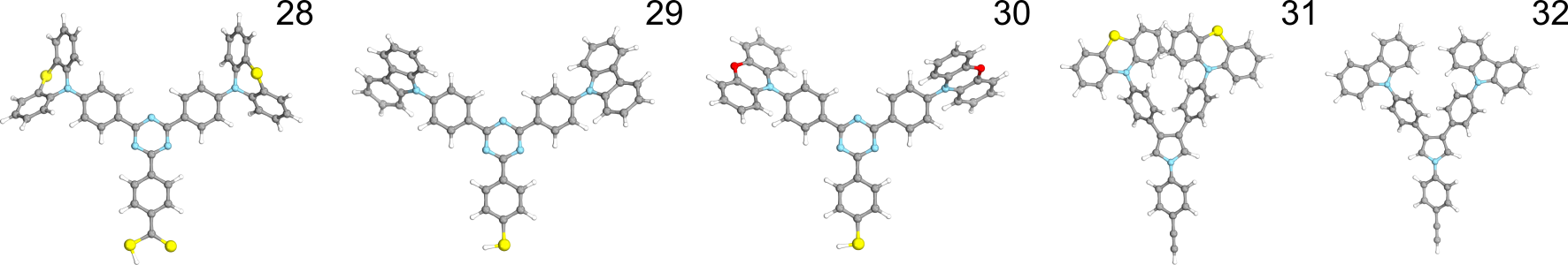}
\caption{Molecular structures of phenyl-substituted derivatives.}
\label{fig:SI8}
\end{figure}

\begin{figure}[h]
\centering
\includegraphics[width=\columnwidth]{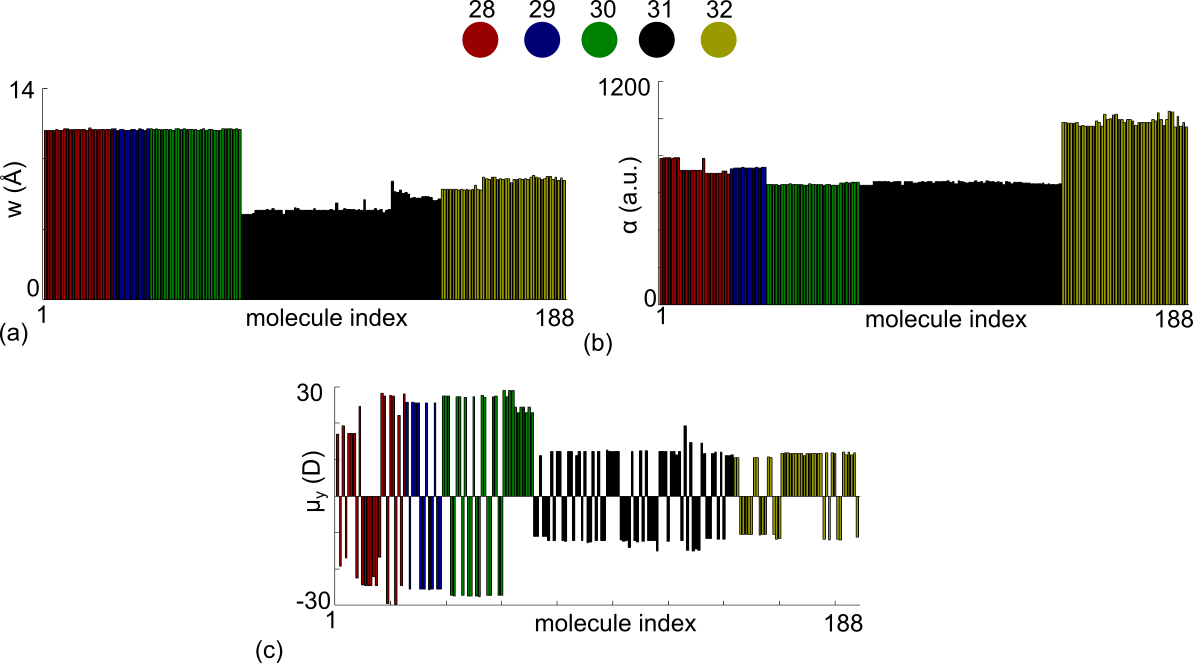}
\caption{Oxidized state properties of phenyl-substituted molecules.}
\label{fig:SI9}
\end{figure}

\begin{figure*}[h]
\centering
\includegraphics[width=\textwidth]{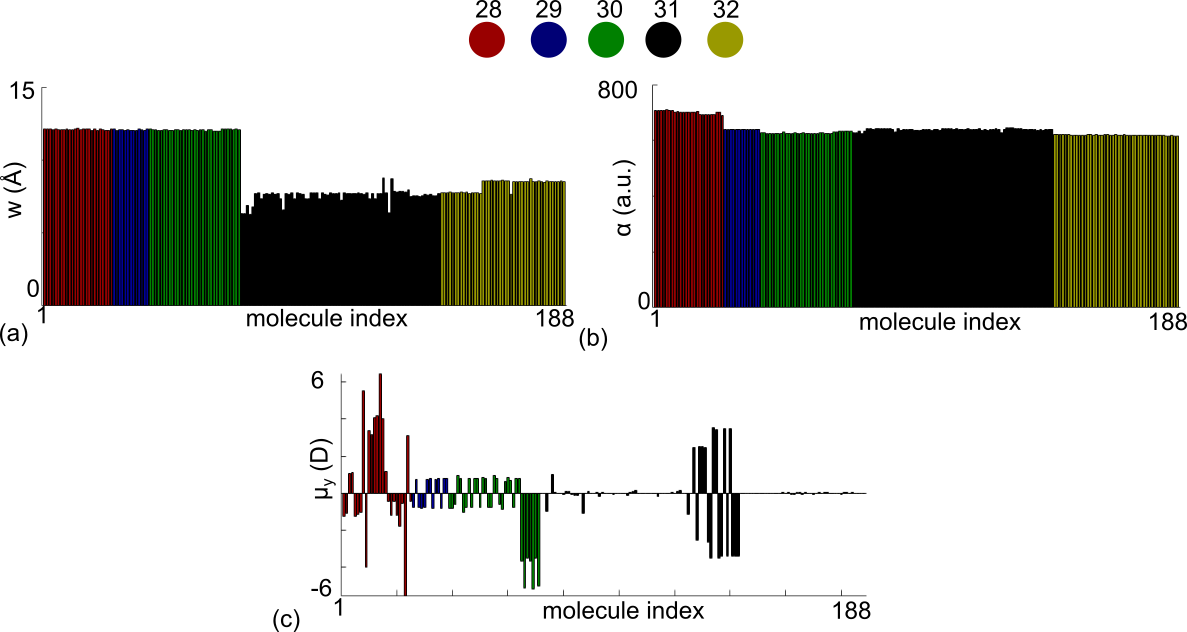}
\caption{Neutral state properties of phenyl-substituted molecules.}
\label{fig:SI10}
\end{figure*}

\clearpage

\section*{5. Semi-Static vs Static Electrostatic Response}
\begin{figure*}[h]
\centering
\includegraphics[width=\textwidth]{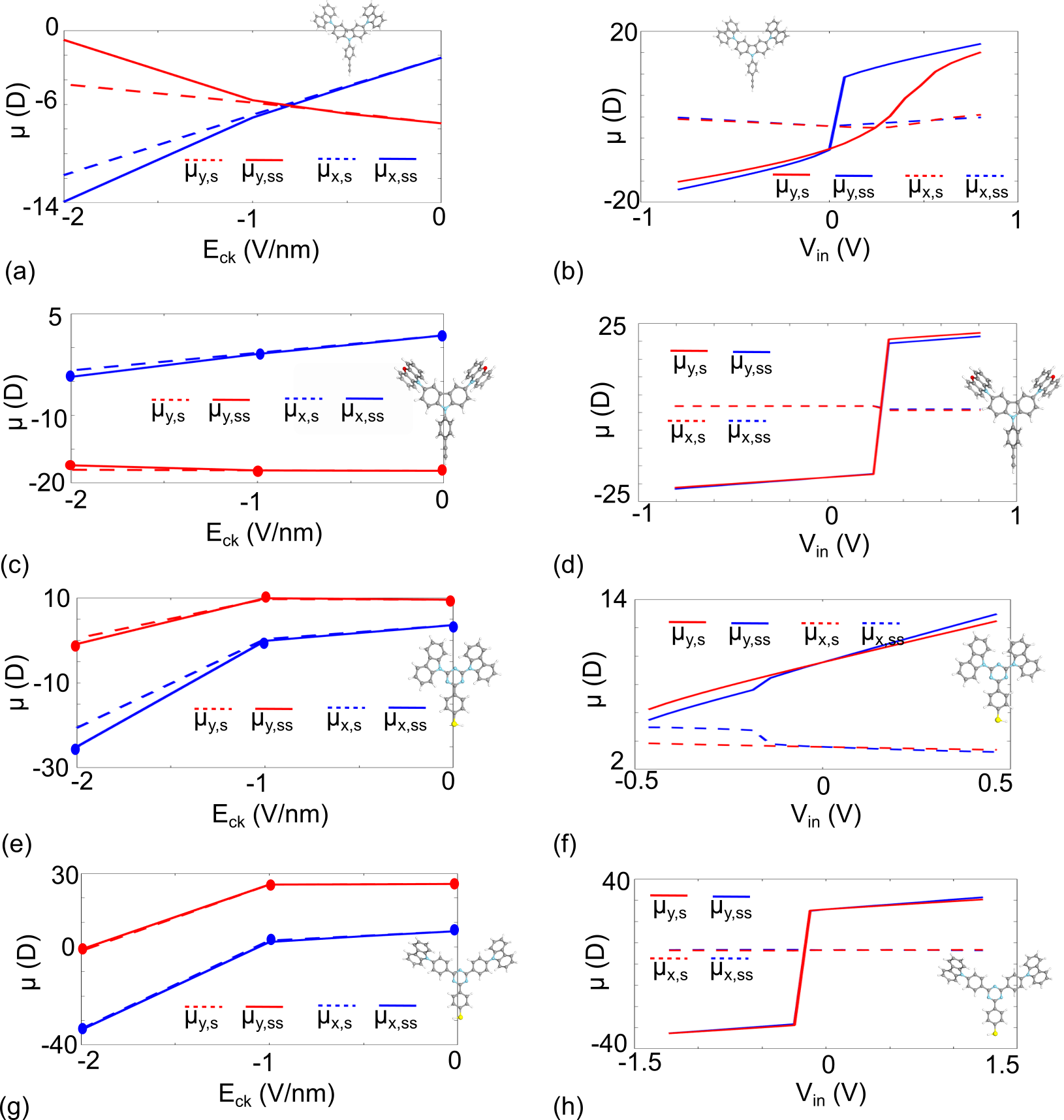}
\caption{Semi-static vs static $V_{in}$–$\mu$ and $E_{ck}$–$\mu$ analysis.}
\label{fig:SI11}
\end{figure*}

\begin{figure*}[h]
\centering
\includegraphics[width=\textwidth]{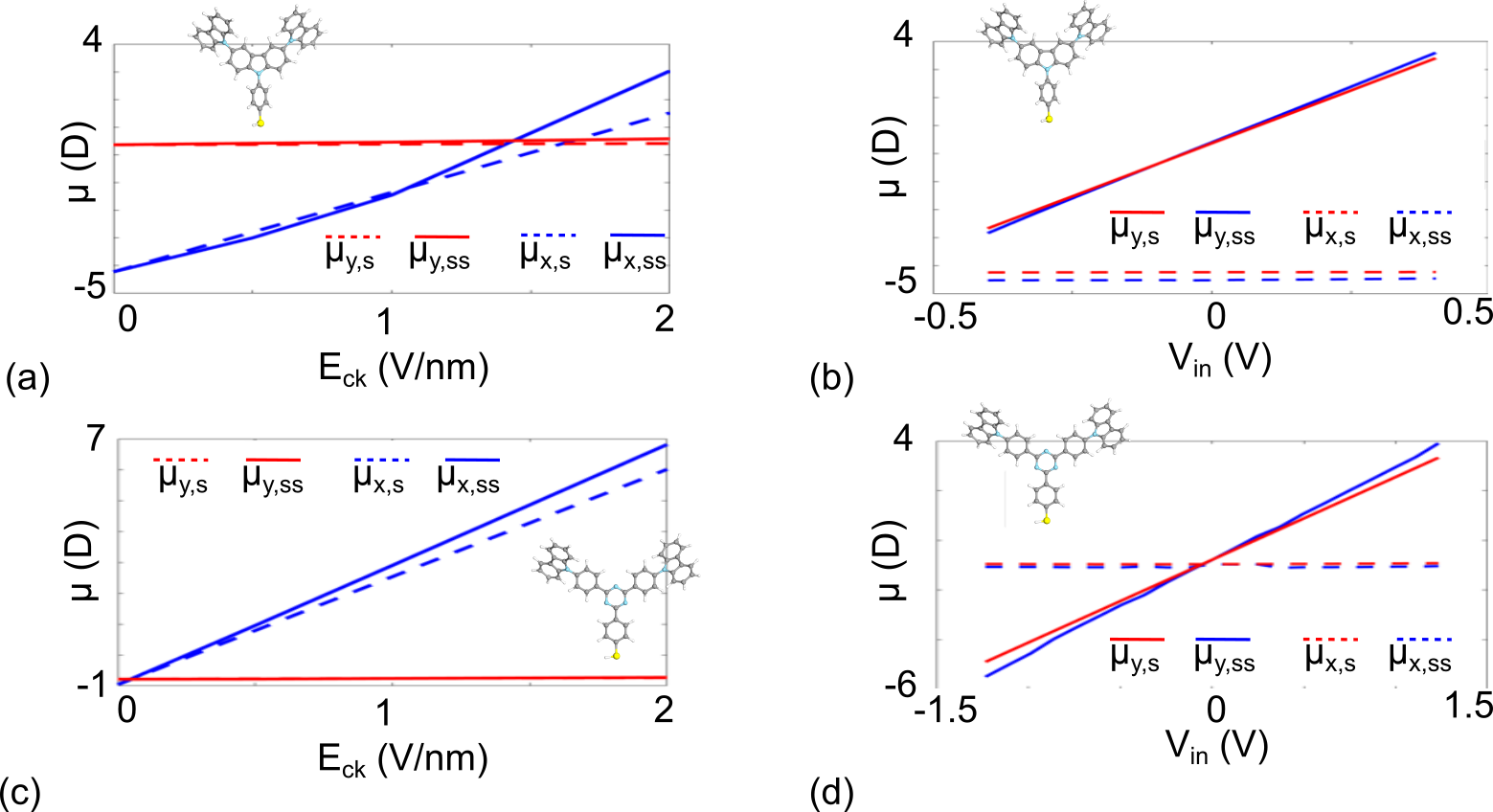}
\caption{Neutral molecules electrostatic response under $E_{ck}$.}
\label{fig:SI11_2}
\end{figure*}

\clearpage

\section*{6. Specularity Coefficient}
To quantify molecular reflection symmetry, we define the specularity coefficient $S$.  
Given atomic coordinates $\mathbf{R} = \{\vec{r}_i\}_{i=1}^N$, the geometric center of mass is:
\[
\vec{r}_{\text{COM}} = \frac{1}{N} \sum_{i=1}^{N} \vec{r}_i
\]

The radius of gyration is:
\[
R_g = \frac{1}{N} \sum_{i=1}^{N} \left\| \vec{r}_i - \vec{r}_{\text{COM}} \right\|
\]

Coordinates are reflected along the Cartesian axes:
\begin{equation}
\begin{aligned}
\vec{r}_i^{(x)} &= (-x_i, \; y_i, \; z_i), \\
\vec{r}_i^{(y)} &= (x_i, \; -y_i, \; z_i), \\
\vec{r}_i^{(z)} &= (x_i, \; y_i, \; -z_i)
\end{aligned}
\end{equation}

The mean deviations are:
\begin{equation}
\begin{aligned}
S_x &= \frac{1}{N} \sum_{i=1}^N \left\| \vec{r}_i - \vec{r}_i^{(x)} \right\|, \\
S_y &= \frac{1}{N} \sum_{i=1}^N \left\| \vec{r}_i - \vec{r}_i^{(y)} \right\|, \\
S_z &= \frac{1}{N} \sum_{i=1}^N \left\| \vec{r}_i - \vec{r}_i^{(z)} \right\|
\end{aligned}
\end{equation}

Normalized specularity coefficients are:
\[
\tilde{S}_x = \frac{S_x}{R_g}, \quad
\tilde{S}_y = \frac{S_y}{R_g}, \quad
\tilde{S}_z = \frac{S_z}{R_g}
\]

Finally, the global specularity coefficient is:
\[
\tilde{S}_{\text{global}} = \frac{1}{3} \left(\tilde{S}_x + \tilde{S}_y + \tilde{S}_z\right)
\]

Lower $\tilde{S}$ values indicate higher specularity. In the main text (Fig.~10c), we report the correlation between $\alpha$ and $\tilde{S}_y$, showing clear separation of Pyr-, Trz-, and Cbz-linked molecules.

\clearpage

\section*{7. Perturbation-Based Grouping Model}
In the aggregated-charge framework, the molecule is mapped onto a reduced three-point charge system, where atomic charges are partitioned into three groups. The resulting model is considered reliable only if it can reproduce the DFT reference values for the dipole moment components ($\mu_x$, $\mu_y$, $\mu_z$) and for the output potential $V_{\text{out}}$ at different $V_{in}$ values under a fixed $E_{ck}$. Conventional spatial-based grouping is, however, too rigid: it does not adapt to changes in the external fields and fails to capture the full electrostatic response of the system.  
To address this issue, we employ a perturbation-based optimization strategy. The method starts from an initial grouping of the atomic charges and evaluates the error between the AC- and DFT-derived $V_{\text{out}}$. Random exchanges of atomic charges between dots 1 and 2 are then performed iteratively on a predefined number of iterations. A new configuration is accepted only if the resulting error decreases compared to the previous step. After a maximum number of iterations, the configuration with the lowest error is retained as the reference for further optimization. This procedure is summarized in Algorithm 1.  
In the subsequent refinement stage, charge exchanges are extended to include dot 3, in order to optimize the out-of-plane contribution ($\mu_x$). At each iteration, the updated configuration is accepted only if it reduces the discrepancy in dipole components while keeping the $V_{\text{out}}$ error below a predefined threshold. This cascade optimization procedure is summarized in Algorithm 2.  

The full perturbation-based algorithm has been applied to all conformers of the molecules considered in this work, ensuring that each conformer is associated with a consistent and optimized set of three point charges. These groupings were then used for VACT extraction.  

\begin{figure*}[t]
\centering
\begin{minipage}[t]{0.48\textwidth}
\begin{algorithm}[H]
\caption{Perturbation method — Step 1 ($V_{\text{out}}$ pre-optimization)}
\begin{algorithmic}[1]
    \State Compute baseline error 
        $error_{\text{last}} \gets \mathrm{Err}(V_{\text{out,AC}}, V_{\text{out,DFT}})$
    \Repeat
        \State Propose new grouping by exchanging two random atomic charges between dots $1$ and $2$
        \State Recompute $V_{\text{out,AC}}$ and set 
        $error_{\text{new}} \gets \mathrm{Err}(V_{\text{out,AC}}, V_{\text{out,DFT}})$
        \If{$error_{\text{new}} < error_{\text{last}}$}
            \State Accept new configuration
            \State $error_{\text{last}} \gets error_{\text{new}}$
        \Else
            \State Reject and revert to previous configuration
        \EndIf
    \Until{maximum iterations reached}
\end{algorithmic}
\end{algorithm}
\end{minipage}\hfill
\begin{minipage}[t]{0.48\textwidth}
\begin{algorithm}[H]
\caption{Perturbation method — Steps 2/3/4 (cascade dipole optimization)}
\begin{algorithmic}[1]
    \State Set current configuration to best from Step~1
    \State Compute baseline dipole error 
        $error_{\text{last}} \gets \mathrm{Err}([\mu_x,\mu_y,\mu_z]_{\text{AC}}, [\mu_x,\mu_y,\mu_z]_{\text{DFT}})$
    \State Define $V_{\text{out}}$ tolerance $\tau_{V_{\text{out}}}$
    \Repeat
        \State Propose new grouping exchanging charges among dots $1$, $2$, $3$
        \State Recompute $([\mu_x,\mu_y,\mu_z]_{\text{AC}}, V_{\text{out,AC}})$
        \State $error_{\text{dip}} \gets \mathrm{Err}([\mu_x,\mu_y,\mu_z]_{\text{AC}}, [\mu_x,\mu_y,\mu_z]_{\text{DFT}})$
        \State $error_V \gets \mathrm{Err}(V_{\text{out,AC}}, V_{\text{out,DFT}})$
        \If{$error_{\text{dip}} < error_{\text{last}}$ \textbf{and} $error_V \le \tau_{V_{\text{out}}}$}
            \State Accept new configuration
            \State $error_{\text{last}} \gets error_{\text{dip}}$
        \Else
            \State Reject and revert to previous configuration
        \EndIf
    \Until{maximum iterations reached}
\end{algorithmic}
\end{algorithm}
\end{minipage}
\end{figure*}

\clearpage

\section*{8. Averaging VACTs}
As discussed in the main text, conformational analysis reveals that the same molecule may present markedly different dipole moments depending on the conformer, leading to different transcharacteristics. Selecting only one conformer for SCERPA simulations would neglect conformational diversity. To address this, we introduce the concept of averaged VACTs, where the effective aggregated charges are computed as the Boltzmann-weighted average over all conformers.

Formally, the average charge for each dot $j$ is:
\begin{equation}
\langle AC_{\text{dot j}} \rangle = \sum_i w_i \cdot AC_{\text{dot j}, i}, \quad j = [1,2,3], , i = [1, N_{\text{conf}}]
\label{eq:Boltz}
\end{equation}

The weights $w_i$ are derived from the single-point energies $E_i$ of each conformer as:
\begin{equation}
w_i = \frac{e^{-\Delta E_i / kT}}{\sum_j e^{-\Delta E_j / kT}}
\label{eq:weights}
\end{equation}

where $\Delta E_i$ is the relative energy of conformer $i$, shifted with respect to the minimum energy conformer:
\begin{equation}
\Delta E_i = E_i - \min_j(E_j)
\label{eq:weight2}
\end{equation}

This approach guarantees that lower-energy conformers dominate the average while higher-energy conformers still contribute proportionally to their thermodynamic relevance. To the best of our knowledge, this represents the first MolFCN modeling explicitly accounting for the role of molecular conformers.

\begin{figure*}
\centering
\includegraphics[width=\textwidth]{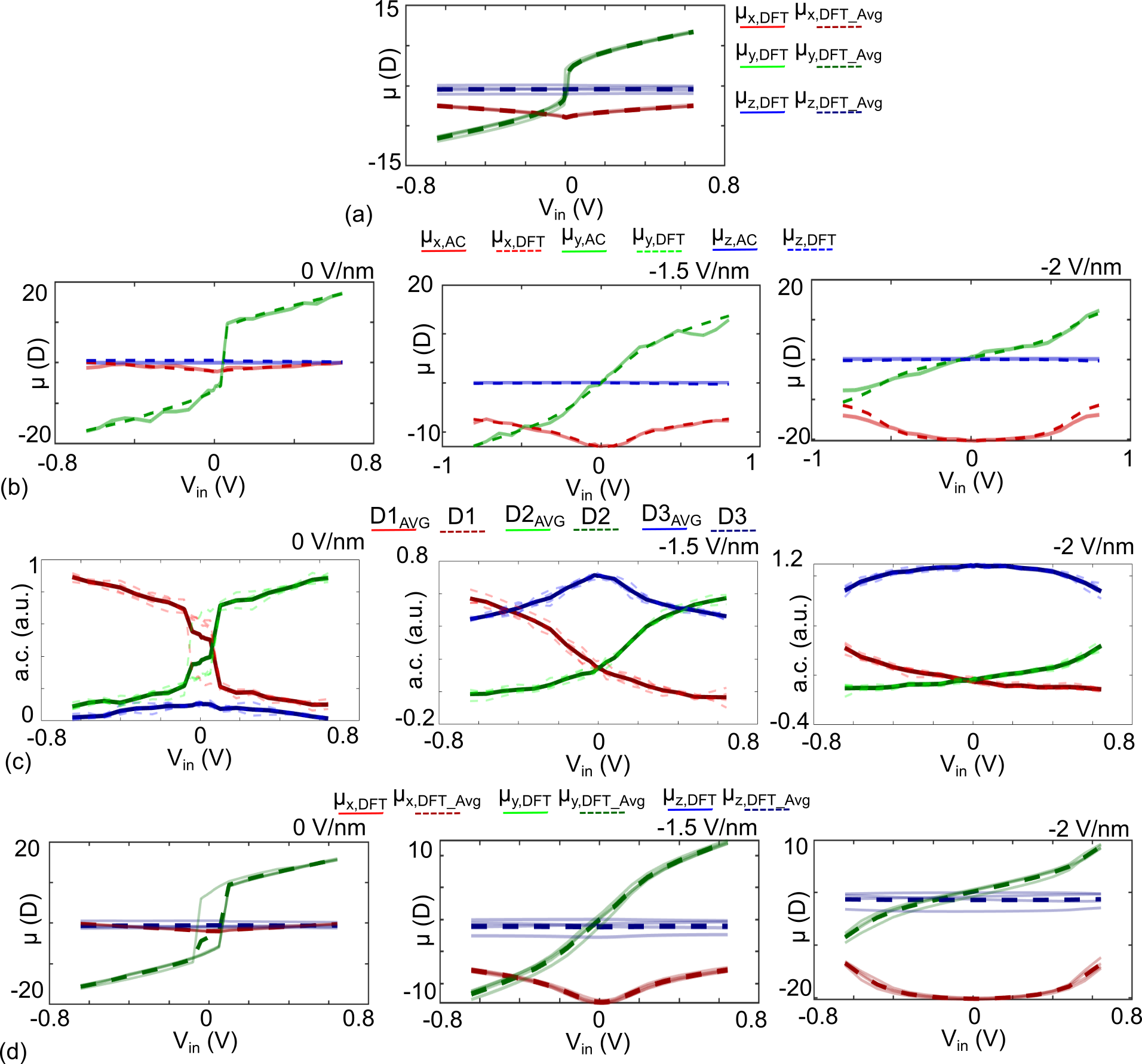}
\caption{Semi-static analysis for molecule \textbf{6} under different clock fields.}
\label{fig:SI12}
\end{figure*}

\begin{figure*}[h]
\centering
\includegraphics[width=\textwidth]{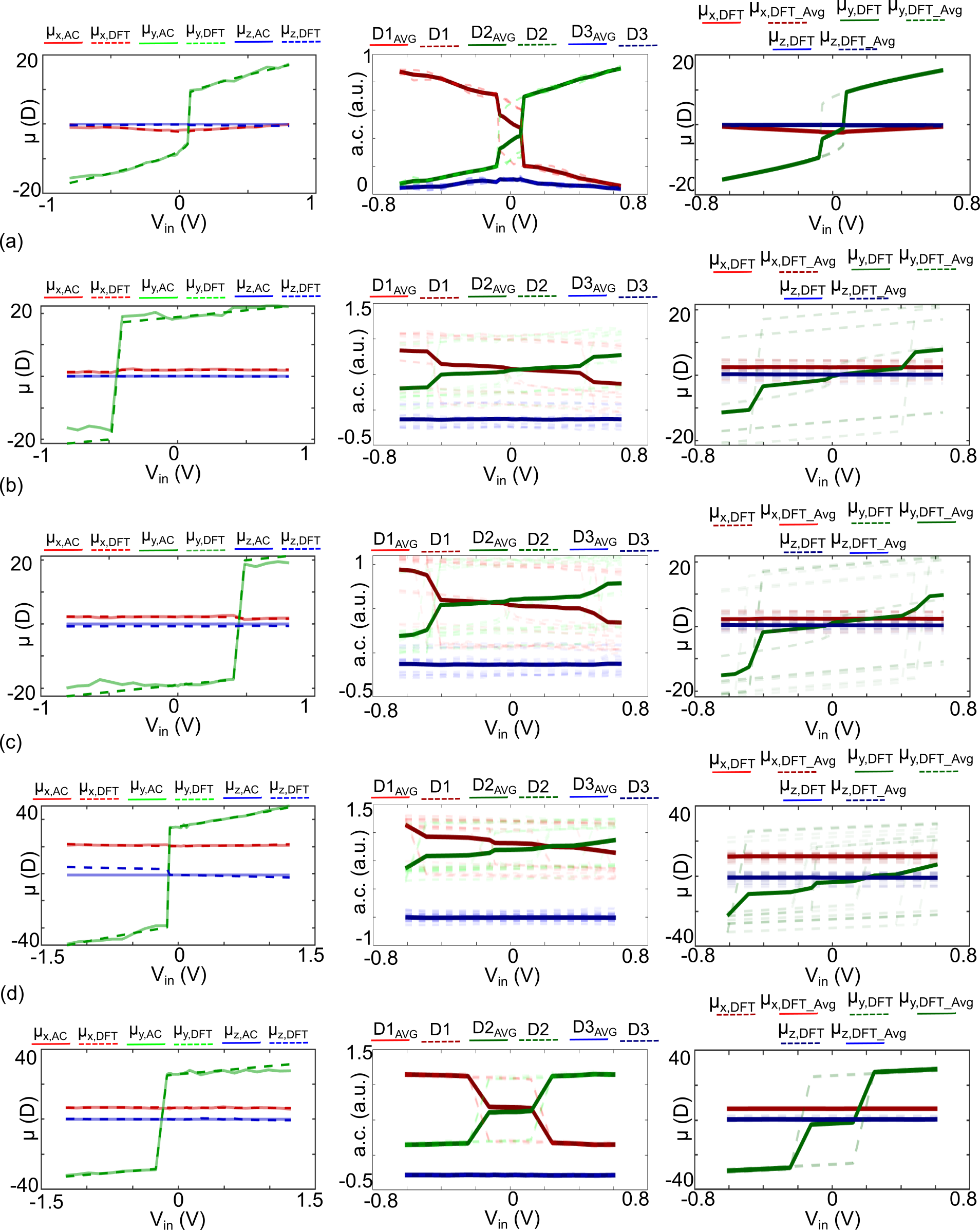}
\caption{Semi-static analysis at $E_{ck}=0$~V/nm.}
\label{fig:SI13}
\end{figure*}

\begin{figure*}[h]
\centering
\includegraphics[width=\textwidth]{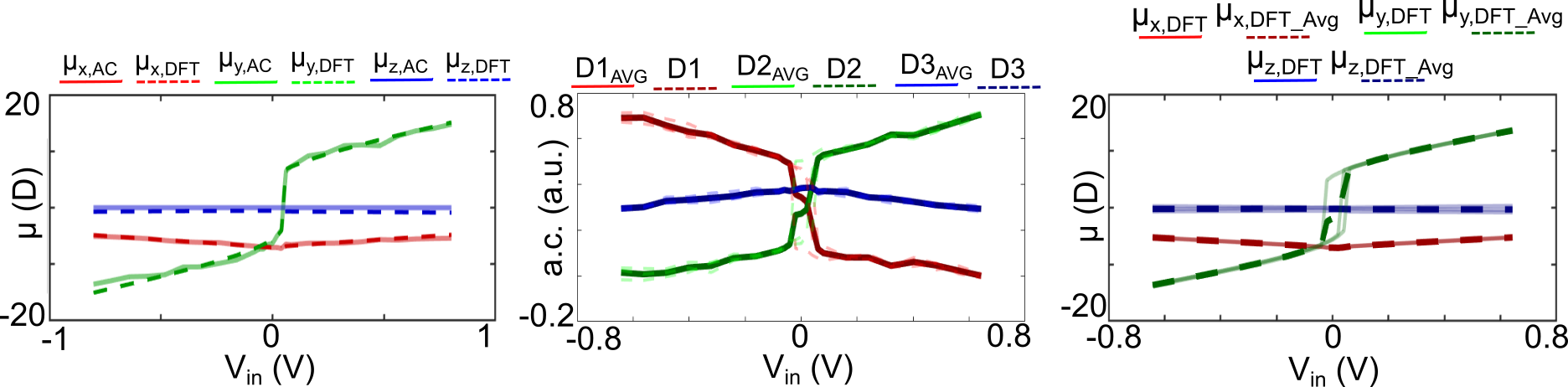}
\caption{Energy-averaged $V_{in}$–$\mu$ curve.}
\label{fig:SI14}
\end{figure*}

\clearpage

\section*{9. AIMD Validation}
\begin{figure*}[h]
\centering
\includegraphics[width=\textwidth]{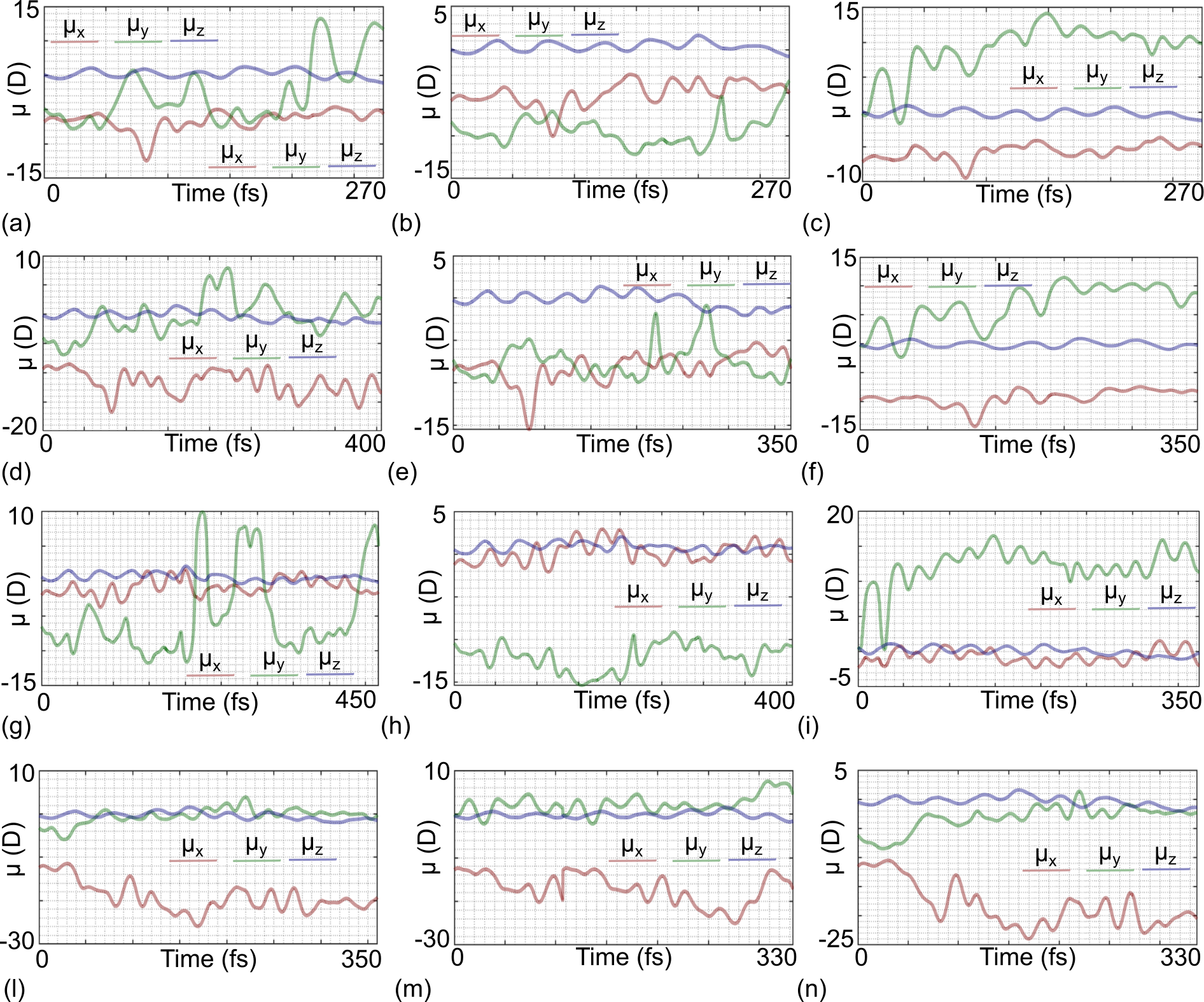}
\caption{Time evolution of dipole moment components in AIMD simulations.}
\label{fig:SI16}
\end{figure*}

\begin{figure*}[h]
\centering
\includegraphics[width=\textwidth]{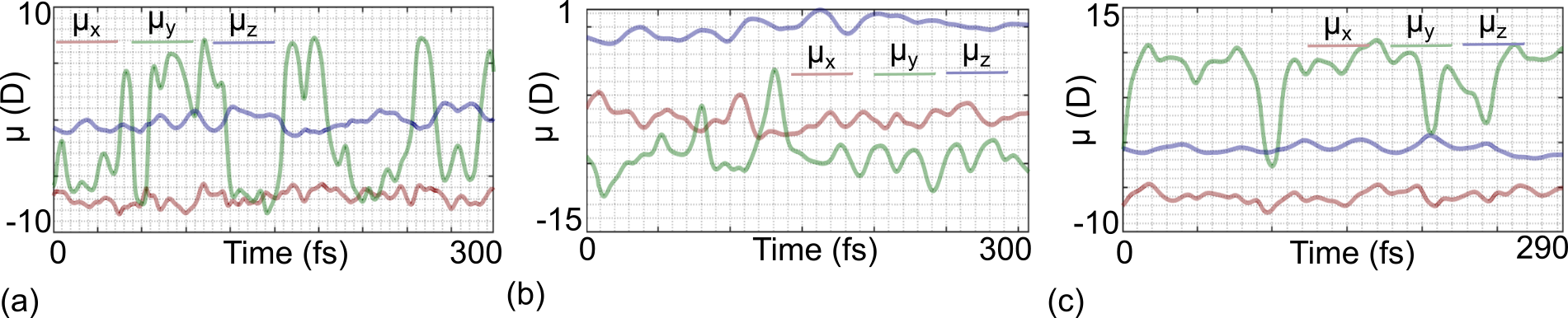}
\caption{Analysis of dipole moment variations for molecule 15.}
\label{fig:SI17}
\end{figure*}

\begin{figure*}[h]
\centering
\includegraphics[width=\textwidth]{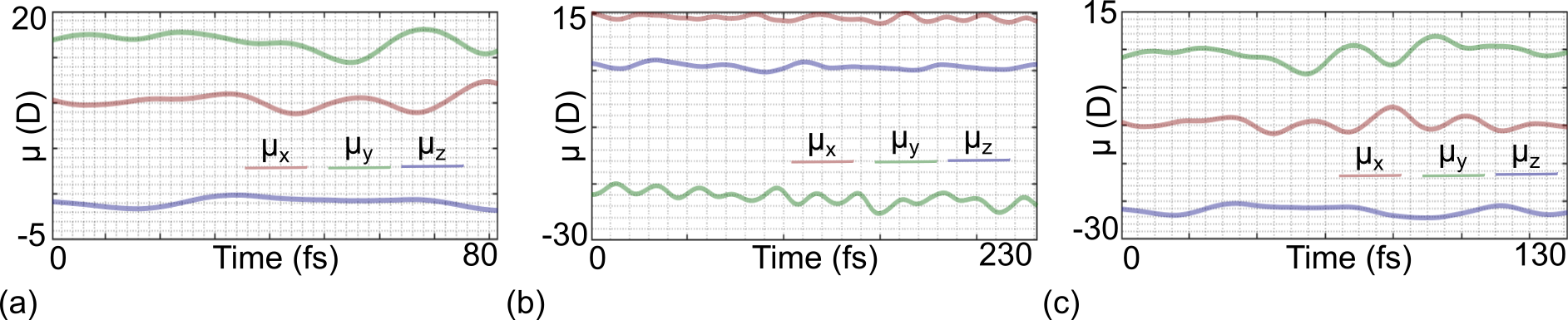}
\caption{Dipole moment dynamics for molecule 28.}
\label{fig:SI18_2}
\end{figure*}

\clearpage

\section*{10. Information Propagation Test}
\begin{figure*}[h]
\centering
\includegraphics[width=\textwidth]{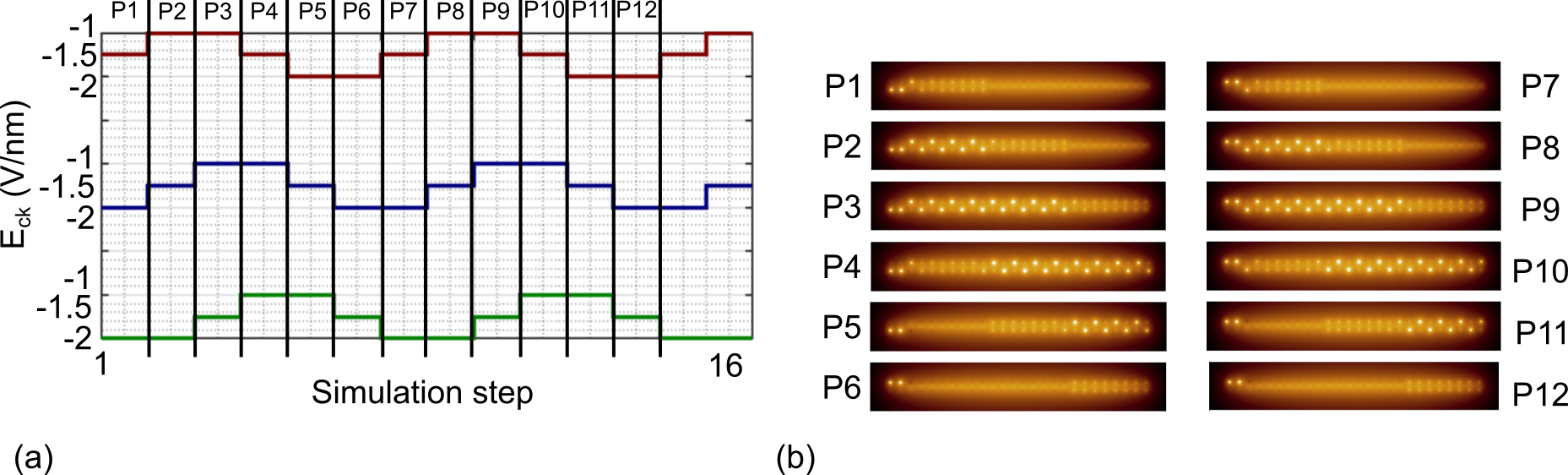}
\caption{Clock signals and information propagation through a molecular wire.}
\label{fig:SI18}
\end{figure*}